\documentclass[sigconf]{acmart}
\usepackage{cleveref}
\usepackage{amsmath}
\usepackage{enumitem}
\usepackage{multirow}
\usepackage{comment}
\usepackage{adjustbox}
\usepackage{cancel}

\usepackage{titlesec}
\titlespacing*{\section}{0pt}{0.5\baselineskip}{0.5\baselineskip}
\titlespacing*{\subsection}{0pt}{0.25\baselineskip}{0.25\baselineskip}

\newcommand{\squishlist}{
 \begin{list}{$\bullet$}
  { \setlength{\itemsep}{0pt}
     \setlength{\parsep}{3pt}
     \setlength{\topsep}{3pt}
     \setlength{\partopsep}{0pt}
     \setlength{\leftmargin}{1em}
     \setlength{\labelwidth}{1em}
     \setlength{\labelsep}{0.5em} } }

\newcommand{\squishend}{
\end{list}  }

\AtBeginDocument{%
  }

\setcopyright{acmcopyright}
\copyrightyear{2025}
\acmYear{2025}
\acmDOI{XXXXXXX.XXXXXXX}

\acmConference[Conference acronym 'XX]{June 03--05, 2018}{Woodstock, NY}

\acmPrice{15.00}
\acmISBN{978-1-4503-XXXX-X/18/06}

\begin{document}

\title{Towards An Efficient LLM Training Paradigm for CTR Prediction}

\author{Allen Lin\textsuperscript{\Large*}}\thanks{{*} Work done during internship at Meta}
\affiliation{%
  \institution{Texas A\&M University}
  \city{College Station, Texas}
  \country{USA}
}

\author{Renqin Cai}
\affiliation{%
  \institution{Meta}
  \city{Menlo Park, California}
  \country{USA}
}

\author{Yun He}
\affiliation{%
  \institution{Meta}
  \city{Menlo Park, California}
  \country{USA}
}

\author{Hanchao Yu}
\affiliation{%
  \institution{Meta}
  \city{Menlo Park, California}
  \country{USA}
}

\author{Jing Qian}
\affiliation{%
  \institution{Meta}
  \city{Menlo Park, California}
  \country{USA}
}

\author{Rui Li}
\affiliation{%
  \institution{Meta}
  \city{Menlo Park, California}
  \country{USA}
}

\author{Qifan Wang}
\affiliation{%
  \institution{Meta}
  \city{Menlo Park, California}
  \country{USA}
}

\author{James Caverlee}
\affiliation{%
  \institution{Texas A\&M University}
  \city{College Station, Texas}
  \country{USA}
}

\begin{abstract}
Large Language Models (LLMs) have demonstrated tremendous potential as the next-generation ranking-based recommendation system. Many recent works have shown that LLMs can significantly outperform conventional click-through-rate (CTR) prediction approaches. Despite such promising results, the computational inefficiency inherent in the current training paradigm makes it particularly challenging to train LLMs for ranking-based recommendation tasks on large datasets. To train LLMs for CTR prediction, most existing studies adopt the prevalent ``sliding-window'' paradigm. Given a sequence of $m$ user interactions, a unique training prompt is constructed for each interaction by designating it as the prediction target along with its preceding $n$ interactions serving as context. In turn, the sliding-window paradigm results in an overall complexity of $O(mn^2)$ that scales linearly with the length of user interactions. Consequently, a direct adoption to train LLMs with such strategy can result in prohibitively high training costs as the length of interactions grows. To alleviate the computational inefficiency, we propose a novel training paradigm, namely \textbf{D}ynamic \textbf{T}arget \textbf{I}solation (DTI), that structurally parallelizes the training of $k$ (where $k >> 1$) target interactions. Instead of constructing a unique training prompt for each target interaction, DTI adopts a streaming prompt formulation strategy to construct one training prompt for $k$ targets and integrates a windowed casual attention mechanism to restrict each target interaction to only attend to its preceding $n$ interaction, resulting in an overall complexity of $O({\frac{m}{k}(n+k)n})$. Furthermore, we identify two major bottlenecks - \textbf{hidden-state leakage} and \textbf{positional bias overfitting} - that limit DTI to only scale up to small value of $k$s (e.g., 5) then propose a computationally light solution to effectively tackle each. Through extensive experiments on three widely adopted public CTR datasets, we empirically show that DTI reduces training time by an average of $\textbf{92\%}$ (e.g., from $70.5$ hrs to $5.31$ hrs), without compromising CTR prediction performance.

\end{abstract}

\maketitle

\section{INTRODUCTION}
\label{sec:intro}
Large Language Models (LLMs) have driven impressive advances in tasks such as context comprehension, reasoning, and modeling world knowledge \cite{zhao2023survey, minaee2024large}. Their remarkable capabilities have fostered growing interest in exploring their potential applications in various domains, including recommendation systems. Many pioneering works \cite{bao2023tallrec, geng2022recommendation, wang2024large} have shown that LLMs can utilize their in-context learning and reasoning abilities to significantly improve the recommendation performance for cold start users or items. Furthermore, with the integration of collaborative filtering knowledge, recent works \cite{bao2023tallrec, zhang2023collm, wu2024coral, kim2024large, lin2024rella} have shown that LLMs can significantly outperform conventional models such as DCNv2 \cite{wang2021dcn}, DIN \cite{zhou2018deep}, and SIM \cite{pi2020search} in ranking-based recommendation tasks, such as click-through-rate (CTR) prediction. 

\begin{figure*}[ht]
  \includegraphics[width=1\textwidth]{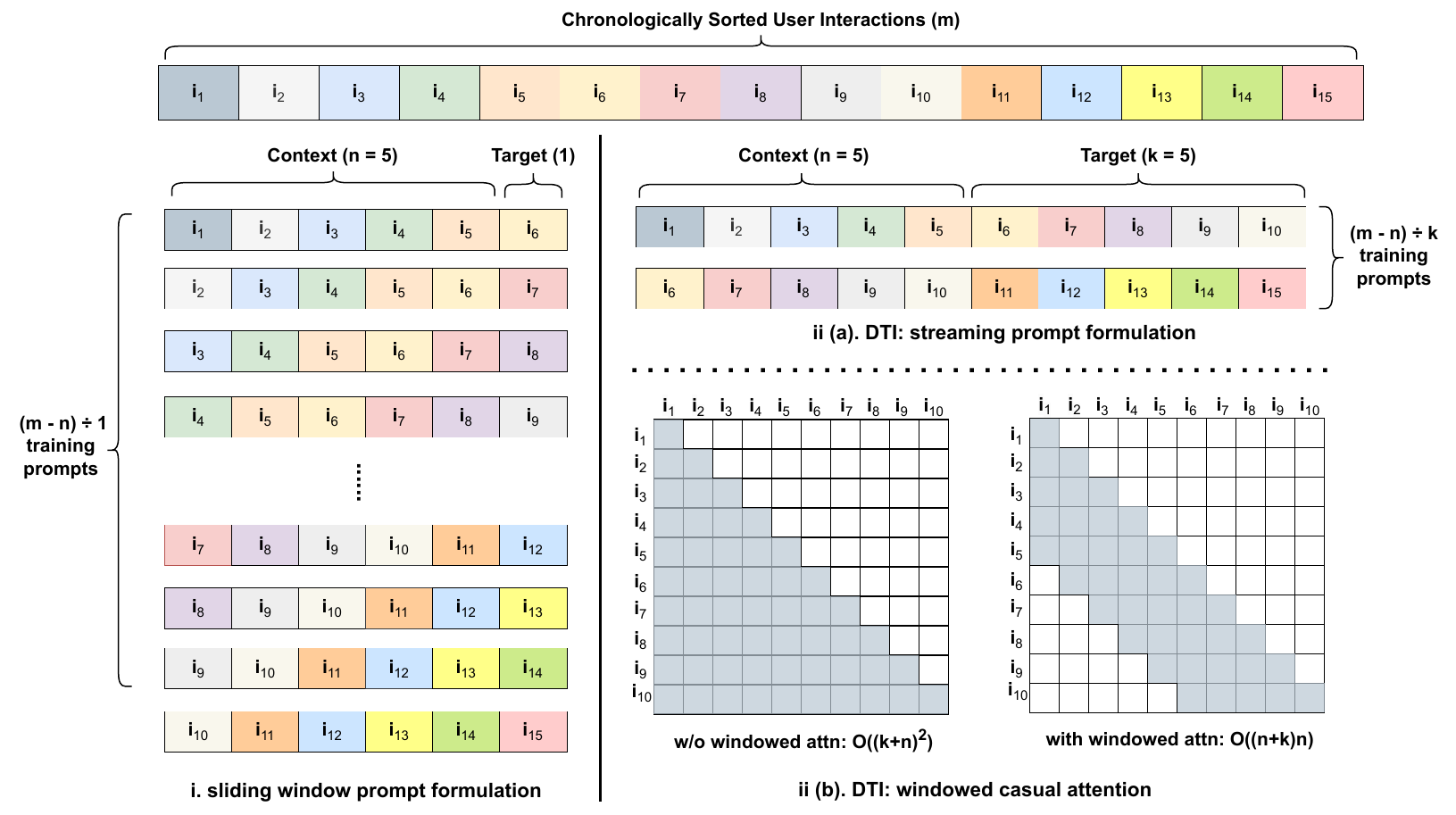}
  \captionsetup{skip=3pt}
  \caption{Overall structural comparison between the sliding-window paradigm and the proposed training paradigm DTI --- (i) illustrates the prompt formulation procedures under the sliding-window paradigm; in contrast, (ii).a illustrates the streaming prompt formulation adopted in DTI. Additionally, (ii).b illustrates the per-streaming-prompt computational complexity both without and with integrating the windowed casual attention mechanism.}
  \label{fig:DTI}
  \vspace{-5pt}
\end{figure*}

In training large language models (LLMs) for click‐through rate (CTR) prediction, recent studies \cite{bao2023tallrec, lin2024rella, geng2022recommendation, zhang2023collm, liao2023llara, kim2024large, wu2024coral, kang2023llms} have adopted the sliding-window paradigm, which is also commonly employed in conventional sequential CTR models \cite{zhou2018deep, qin2020user, kang2018self}. Specifically, given a chronologically ordered sequence of $m$ user interactions, a unique prompt is constructed for each interaction by designating it as the prediction target and using its $n$ preceding interactions as context, resulting in a total of $m-n$ prompts. Consequently, training on a target interaction requires processing all its $n$ context interactions, resulting in a computational cost that scales quadratically with the context size (i.e., $O(n^2)$ per prompt). Summed over all $m-n$ prompts, this leads to an overall complexity of $O(mn^2)$. Different from conventional sequential CTR models that represent each interaction with one or a few id tokens, LLMs represent each interaction using its textualized descriptions, leading to a much longer context. Take the MovieLens dataset as an example, a context of 20 preceding interactions takes only $\textbf{20}$ tokens (one feature id per interaction) for SASRec \cite{kang2018self} to represent but can take up to $\textbf{701}$ tokens for Llama 3 \cite{dubey2024llama} -- see a sample prompt in the Appendix (\Cref{fig:prompt}). This expanded context length magnifies computational complexity by more than $800$ times under the $O(mn^2)$ scaling, making directly adopting the sliding-window paradigm for LLM CTR training computationally challenging. 

To address this challenge, we propose a novel \textbf{D}ynamic \textbf{T}arget \textbf{I}solation (DTI) paradigm that integrates a streaming prompt formulation strategy to structurally parallelize the training of $k$ (where $k >> 1$) target interactions (items). As shown in \Cref{fig:DTI}, instead of constructing a unique prompt for each target interaction, DTI constructs one prompt for $k$ targets by appending $k$ consecutive target interactions after the first $n$ context interactions to form an extended prompt comprising $n + k$ interactions. To train on the first target interaction, the LLM encodes the hidden state of all previous $n$ context interactions leading to it, just as in the sliding-window paradigm; however, for each subsequent target interaction, all its preceding interactions can serve as the context without the need to redundantly re-encode them in separate forward passes. Consequently, the streaming prompt formulation can be considered as traversing the user interaction sequence in a stride of $k$, reducing total constructed prompts from $\mathbf{m-n}$ in the sliding‐window paradigm to $\mathbf{\frac{m}{k}}$. 

Next, a windowed casual attention mechanism is integrated to restrict each target interaction to only attend to its preceding $n$ interactions during training. This mechanism reduces the per-prompt complexity from $O((n+k)^2)$ (since each prompt now has $n + k$ interactions instead of $n$) to $O((n+k)n)$. Together, DTI reduces the overall complexity by a factor of $\mathbf{k}$, from $\mathbf{O(mn^2)}$ in the sliding-window approach to $O(\mathbf{\frac{m}{k}(n+k)n})$ -- yielding an average training time reduction of $\mathbf{92\%}$ when empirically tested on three widely adopted datasets - \Cref{tab:DTI_time_red}. 

While DTI dramatically increases training efficiency, when empirically tested on three widely adopted datasets, a consistent performance decrease is observed as the value of $k$ increases as shown in \Cref{fig:DTI_base}. We identify the two root causes -- \textbf{hidden-state leakage} and \textbf{positional bias overfitting} -- that constrain DTI to only scale to small values of $k$ (e.g., 5) and propose a computationally light solution for each. Specifically, even with windowed casual attention, a target interaction can still peak into prior interaction beyond its intended context window, since its context interactions can also attend to their preceding $n$ interaction, namely, \textbf{ hidden-state leakage}. This causes the LLM to learn to make CTR predictions based on a prolonged context, creating a discrepancy between training and inference. To address this, we propose a computationally-light yet effective hidden-state resetting mechanism to partially interpolate the final hidden state of each context interaction to mitigate the leakage. In addition to hidden-state leakage, since each training prompt now contains $k$ targets, the LLM could inadvertently learn to model the positional encoding of each target when making CTR predictions, leading to \textbf{positional bias overfitting}. To address this, we eliminate the use of absolute position ids for each target but rather directly encode its positional information via its relative distance between preceding tokens. This forces the model to base its predictions on the semantic dependencies between each target and its corresponding context, rather than its absolute position id that encourages incorrect positional patterns. In sum, the main contributions of this work are as follows:
\begin{itemize}[topsep=5pt, partopsep=5pt, itemsep=1pt]
\item We propose a novel \textbf{D}ynamic \textbf{T}arget \textbf{I}solation (DTI) paradigm that incorporates a streaming prompt formulation and a windowed casual attention mechanism to structurally parallelize the training of $k$ target interactions, significantly increasing the computational efficiency compared to the conventional sliding-window paradigm. 
\item We empirically and theoretically demonstrate two major bottlenecks \textbf{--} \textbf{hidden-state leakage} and \textbf{positional bias overfitting} \textbf{--} that restrict DTI to only scale to small $k$s (e.g., 5) before significantly degrading in performance; and propose a computationally light solution that effectively addresses each.
\item Through extensive experiments on three widely adopted datasets, we demonstrate that the proposed training paradigm, DTI, reduces the LLM's training time until convergence by an average of \textbf{92\%} compared to the conventional sliding-window approach, while incurring minimal to no trade-offs in performance or inference efficiency.
\end{itemize}

\section{PRELIMINARIES}
\label{sec:pre}
\textbf{a. Notations.} Let $\mathcal{U}$ and $\mathcal{I}$ denote the set of users and items, respectively. For a given user $u \in \mathcal{U}$, $S^{u} = (i_1^u, i_2^u, ..., i_m^u)$ is a sequence of $m$ item interactions of a user $u$ sorted in a chronological order, where $i_j^u \in \mathcal{I}$ represents the $j$-th item the user has engaged with. Each item $i_j^u$ also has a corresponding label $l_j^u$ that reflects whether the user enjoyed the interaction or not (yes or no).

\noindent\textbf{b. LLM for Click Through Rate Prediction.} We study the application of decoder-only LLMs (due to their overwhelming prevalence) to the classic sequential (user-behavior-based) click-through rate prediction task. Essentially, the LLM aims to predict the label that the user will give to an arbitrary target interaction, based on a sequence of past interactions used as the context. To achieve this, existing works \cite{bao2023tallrec, lin2024rella, geng2022recommendation, zhang2023collm, liao2023llara, kim2024large, wu2024coral, kang2023llms} adopt the commonly used sliding-window paradigm to formulate training prompts for each target interaction by using its preceding $n$ interactions as the context. To effectively utilize LLM's pretraining natural language understanding, both the target and the context interactions are being represented using their item descriptions -- see a sample prompt in Appendix (\Cref{fig:prompt}). Consequently, the resulting prompt is much longer than the id-based model inputs for conventional user-behavior based CTR models, further exacerbating the aforementioned computational redundancy of the sliding-window paradigm. To alleviate the computational burden, we propose a novel \textbf{D}ynamic \textbf{T}arget \textbf{I}solation (DTI) paradigm that parallelizes the training of $k$ (where $k >> 1$) target interactions in one single prompt.

\noindent\textbf{c. Pointwise Scoring with LLMs.} Since CTR prediction requires the LLM to perform pointwise scoring to output a floating point value between 0 and 1, we follow previous works \cite{lin2024rella} to conduct bi-dimensional softmax over the predicted probabilities of the 'yes' and 'no' token.

\section{PROPOSED TRAINING PARADIGM - DTI}
In this section, we introduce the details of the proposed \textbf{D}ynamic \textbf{T}arget \textbf{I}solation (DTI) paradigm, and propose a method to quantify the reduction of training FLOPs in comparison with the conventional sliding-window paradigm.

\subsection{The Sliding Window Approach}
To provide a better understanding of DTI, we first introduce the computational redundancy inherent in the widely adopted sliding-window paradigm. Given a sequence of chronologically sorted user interactions $S^{u} = (i_1^u, i_2^u, ..., i_m^u)$, a unique training prompt is constructed for each interaction starting from the $i_{n+1}^{u}$ interaction by using its preceding $n$ interactions as context. Each prompt comprises $n$ context interactions followed by one target interaction, with the objective of teaching the LLM to predict whether the user will favor the target interaction based on the user's most recent $n$ interactions. In turn, the sliding-window approach can be considered as traversing the sorted sequence of user interactions using a window of size $n$ with a stride of one. For a sequence of $\mathbf{m}$ user interactions, it generates a total of $\mathbf{m - n}$ \textbf{distinct training prompts}. Although this training prompt formulation strategy is prevalent in the literature \cite{bao2023tallrec, lin2024rella, geng2022recommendation, zhang2023collm, liao2023llara, kim2024large, wu2024coral, kang2023llms}, it incurs \textbf{a substantial amount of redundant computation}. Due to the sequential nature of the interaction sequence, any two constructed training prompts for target interactions within a distance $d$ (where $d < n$) of each other will share $n-d$ overlapping context interactions. For instance, the training prompts constructed for target interaction $i_{n+1}^{u}$ and target interaction $i_{n+2}^{u}$ respectively possesses context sequences $(i_{1}^{u}, i_{2}^{u}, ... i_{n}^{u})$ and $(i_{1+1}^{u}, i_{2}^{u}, ... i_{n+1}^{u})$, sharing a total of $n-1$ common context interactions. When these two prompts are processed independently during training, the LLM redundantly computes the hidden state for the $n - 1$ overlapping context interactions for each target, leading to a computational burden that increases with the length of the user sequence. 

\subsection{Streaming Prompt Formulation}
\label{subsec:prompt_formulation}
Inspired by the sequential nature of the interaction data, DTI leverages a streaming prompt formulation strategy that reuses prior target interactions as contextual inputs for subsequent target predictions, thereby dramatically mitigating redundant computation. Specifically, rather than constructing isolated prompts for each target interaction — as in the sliding-window paradigm — DTI appends $k$ consecutive target interactions to the initial $n$ context interactions, resulting in a prompt with a total $n + k$ interactions. In this configuration, each subsequent target interaction not only benefits from the initial $n$ context interactions but can also freely reuse the hidden states of preceding target interactions within the same prompt. Consequently, once a target interaction’s hidden state is computed, it is immediately available to serve as context for predicting subsequent target interactions, eliminating the need for re-computing the hidden states of highly overlapping context in separate forward passes, as in the sliding-window approach. Ultimately, the streaming prompt formulation can be considered as traversing the sorted sequence of user interactions using a window of size $n$ but with a stride of $k$. Given the same sequence of $m$ user interactions, DTI generates a total of  $\mathbf{\frac{m}{k}}$ \textbf{training prompts}, compare to $\mathbf{m - n}$ \textbf{training prompts} under the conventional sliding-window paradigm -- as shown in \Cref{fig:DTI}.ii(a). 

Although the stream prompt formulation strategy alleviates computational redundancy by reducing constructed training prompts by a factor of $k$, it encompasses two constraints. First, appending $k$ target interactions makes the resulting training prompt $k-1$ interactions longer than that produced by the sliding-window approach, which appends only one target interaction after the context interactions. This increased prompt length can cause significant computational and memory overhead due to the quadratic scaling of the self-attention mechanism in LLMs, especially when a large value of $k$ is chosen. This overhead may substantially offset the computational efficiency gains if not properly addressed. Secondly, training the LLM with the streaming prompt formulation creates a dilemma during inference. During training, a target interaction leverages all preceding interactions as context; specifically, the $j$th (where $j$ ranges from $1$ to $k$) target interaction uses a context comprising $n+j-1$ interactions. Consequently, for target interactions appearing later in the prompt, the LLM may inadvertently learn to base CTR predictions on a much longer context than the intended $n$ preceding interactions. This discrepancy poses a challenge during inference: the LLM, having been frequently trained with an extended context, expects a similar window length during prediction to achieve optimal performance. However, expanding the context window beyond $n$ interactions during inference results in significantly increased latency due to quadratic scaling, which can be detrimental in many applications.

\subsection{Windowed Casual Attention}
To address the two aforementioned constraints, we propose a windowed causal attention mechanism to be integrated into DTI. During training, DTI restricts each target interaction to attend only to its preceding $n$ interactions. Formally, the windowed casual attention is defined as follows:
\begin{equation}
    \text{Attention}_{\text{window}}(Q, K, V)_t = \sum_{s=\max(1, \, t - N)}^{t} \alpha_{t,s} \, V_s \\
\end{equation}

\begin{equation}
    \alpha_{t,s} = \frac{\exp\left(\frac{Q_t \cdot K_s^\top}{\sqrt{d_k}}\right)}{\sum_{k=\max(1, \, t - N)}^{t} \exp\left(\frac{Q_t \cdot K_k^\top}{\sqrt{d_k}}\right)},
\end{equation}

\noindent where $Q, K, V$ denotes the Query, Key, and Value matrices, and $\alpha_{t,s}$ denotes the attention score that the target interaction $t$ assigns to an arbitrary token $s$ in the causal attention window of size $N$. This straightforward modification yields two major benefits. First, the windowed causal attention mechanism reduces the computational complexity of a streaming prompt from $O((k+n)^2)$ to $O((n+k)n)$, as shown in \Cref{fig:DTI}.ii(b). Since each target interaction now attends only to a fixed subset of $n$ preceding interactions, both the computational and memory complexities scale linearly with the increased prompt length rather than quadratically. Second, it directly resolves the inference dilemma introduced by the streaming prompt formulation. By restricting each target interaction to attend solely to its preceding $n$ interactions, the LLM is trained to base its predictions on a uniformly sized context window, thereby aligning training and inference procedures. This improvement allows for the selection of a large $k$ value without incurring the computational burdens associated with quadratic scaling, making the adoption of the streaming prompt formulation more meaningful.

\subsection{Training Setting}
To integrate the streaming prompt formulation and the windowed causal attention during training, we insert $k$ special [SUM] tokens throughout the prompt, one immediately after each target interaction -- see a sample streaming prompt in Appendix (\Cref{fig:stream_prompt}). Each [SUM] token is designated to attend to its corresponding target interaction along with that target's $n$ preceding context interactions, thereby producing a hidden state summary that encapsulates the likelihood of the target being favored by the user. During training, DTI computes the language modeling cross-entropy loss between the final hidden states of the $k$ [SUM] tokens and their corresponding textual ground-truth labels ('yes' or 'no'). This loss is then averaged and backpropagated to update the parameters of the LLM. Formally, the training objective is defined as:
\begin{equation*}
    \max_{\Theta} \sum_{(x,y) \in \mathcal{M}} 
    \sum_{j=1}^{k}
    \log 
    P_{\Theta}\Bigl(
      y_{j}^{m}
      \;\Bigm|\;
      x^{m}_{\max\{1,\,i_{j}^{m} - N\}}
      :\,
      x^{m}_{\,i_{j}^{m}}
    \Bigr),
\end{equation*}

\noindent where $M$ denotes the set of $\frac{m}{k}$  training prompts constructed using the streaming prompt formulation approach; $k$ denotes the total number of target interactions / [SUM] tokens in each prompt; $y_j$ denotes the $j$-th label for the $j$-th [SUM] token; and $i_j$ denotes the position of the $j$-th [SUM] token in the prompt. The \textbf{windowed casual attention} mechanism:
\begin{equation*}
    x_{\max\{1,\,i_{j} - N\}}: x_{\,i_{j}}
\end{equation*}
is applied such that each [SUM] token can only attend to tokens associated with its preceding $n$ context interactions.\footnote{For clarity, we use $N$ to denote the size of the causal attention window, which corresponds to the number of tokens associated with the $n$ context interactions}

\subsection{Training FLOPs Reduction}
We formally quantify the training FLOPs to measure the amount of computation that DTI reduces in comparison with the conventional sliding-window paradigm. Specifically, we compute the FLOPs incurred by each paradigm on a sequence of $m$ chronologically sorted user interactions via $n$ preceding interactions serving as the context. Since the sliding-window formulation results in a total of $m - n$ distinct training prompts, the total training FLOPs can be approximated as follows:
\begin{equation*}
    \text{Sliding Window FLOPs} \approx (m-n) \times 2L \times \left( N^2 d + N d^2 \right),
    \label{eq:flops_basic}
\end{equation*}
where L denotes the number of layers in the LLM (multiplied by 2 due to both the forward and the backward pass), d denotes the hidden dimension size, and N denotes the padded length of each training prompt. Next, we compute the total training FLOPs of DTI. Since the streaming prompt formulation only results in $\frac{m}{k}$ training prompts, DTI's total FLOPs can be approximated as follows:
\begin{equation*}
    \text{DTI FLOPs} \approx \frac{m}{k} \times 2L \times \left( (N + K) \times N \times d + (N + K) \times d^2 \right),
    \label{eq:flops_limited}
\end{equation*}
where $N+K$ denotes the length of the extended prompt, d denotes the hidden dimension size, and N denotes the size of the casual attention window. The training FLOPs reduction switching from the sliding-window paradigm to DTI can then be quantified as:
\begin{align*}
    \text{FLOPs Reduction} &= \frac{2L \cdot (N^2 \cdot d + N \cdot d^2) \cdot (m - n)}{2L \cdot \left[(N + K) \cdot N \cdot d + (N + K) \cdot d^2\right] \cdot \left(\frac{m}{k}\right)} \\
    &= \frac{2L \cdot N \cdot d \cdot (N + d) \cdot (M - N)}{2L \cdot (N + K) \cdot d \cdot (N + d) \cdot \left(\frac{m}{k}\right)} \\
    &= \frac{\cancel{2L} \cdot {N} \cdot \cancel{d} \cdot \cancel{(N + d)} \cdot (m - n)}{\cancel{2L} \cdot (N + K) \cdot \cancel{d} \cdot \cancel{(N + d)} \cdot \left(\frac{m}{k}\right)} \\
    &= \frac{N \cdot (m - n)}{(N + K) \cdot \left(\frac{m}{k}\right)} = \frac{N \cdot (m - n) \cdot k}{(N + K) \cdot m} \\
    &= \frac{N \cdot k (m - n)}{m (N + K)}
\end{align*}

In reality, the user sequence $m$ is often much larger than the number of context interactions used $n$, therefore, we can approximate $(m-n)$ down to $m$ and simplify the above equation to:

\begin{equation}
    \text{FLOPs Reduction} \approx \frac{N \cdot k}{(N + K)},
    \label{eq:flop_red}
\end{equation}

\noindent where $k$ denotes the number of target interactions included in one streaming prompt; and $N$ and $N + K$ respectively denotes the length of a sliding-window prompt and a streaming prompt. To provide a concrete example, Let the sliding-window paradigm construct each prompt using the preceding $20$ interactions as the context by the target interaction ($n=20$), and the streaming prompt formulation to include $50$ targets in each prompt ($k=50$). Then, ${\frac{N}{(N+K)}}$ boils down to $\frac{20c}{(20+50)c}$, where $c$ denotes the number of tokens needed to represented an interaction. If we further substitute in $50$ for $k$, we get an estimated FLOPs reduction of $\mathbf{14.28}$ times compared to the sliding-window paradigm. Intuitively, increasing the number of targets per-streaming-prompt (i.e. a larger $k$) leads to more training FLOPs reduction. 

\subsection{Inference Setting}
During inference, we follow the conventional sliding-window paradigm to form a unique prompt for each test target interaction, using its previous $n$ interactions as context. This is chosen both to make a fair comparison with the current training paradigm, and, more importantly, to guarantee inference efficiency. It is important to note that DTI adopts two different prompt formulation strategies during training and inference; however, the integrated widowed casual attention mechanism greatly reduces the discrepancy between training and inference. To further mimic what the LLM sees during training, we append a [SUM] token at the end of each test prompt, and use its final hidden state to derive the logits of 'yes' and 'no' to compute the AUC and the F1 score as introduced in \Cref{sec:pre}c.

\begin{figure}
  \includegraphics[width=.49\textwidth]{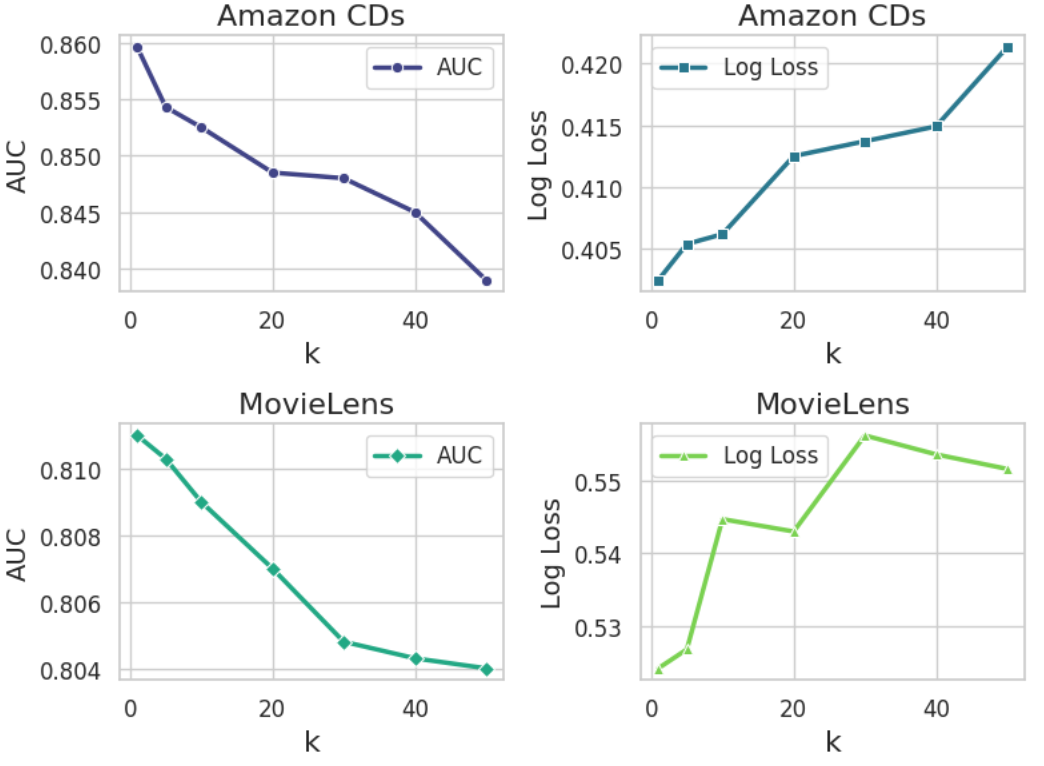}
  \captionsetup{skip=2pt}
  \caption{CTR prediction evaluations (AUC$\uparrow$ and Log Los$\downarrow$) of DTI with different numbers of target interaction ($k$) per streaming prompt.}
  \label{fig:DTI_base}
\end{figure}

\section{DTI Bottlenecks and Solutions}
Although the proposed DTI training paradigm significantly reduces the training cost, as shown in \Cref{fig:DTI_base}, the CTR prediction performance of trained LLMs decreased steadily as the value of $k$ increases, when empirically tested on two widely adopted datasets. In this section, we identify two root causes -- \textbf{hidden state leakage} and \textbf{positional bias overfitting} -- that constrain DTI from scaling to larger values of $k$ while maintaining the same level of CTR performance, and propose a tailored solution for each.

\subsection{Hidden State Leakage}
\smallskip
\noindent\textbf{Cause.}
Under the streaming prompt formulation, each subsequent target interaction could freely use all preceding interactions as context without the need to re-encode their hidden states in separate forward passes. This elimination of redundant computations is the core training advantage that contributes to DTI's FLOPs reduction. However, even when combined with the windowed causal attention mechanism, the reuse of these static hidden states of previous interactions introduces a critical discrepancy between training and inference, namely the hidden state leakage. During training, since DTI adopts the streaming prompt formulation, the first layer hidden state $\textbf{h}_t^1$ of an arbitrary target interaction $t$ can be defined as follows: 
\vspace{-0.1cm}
\begin{equation}
    \label{eq:h_1}
    \textbf{Layer 1:}
    \quad
    \textbf{h}_t^1 = \sum_{j = t-N}^{t} \alpha_{t,j}^1 \,\bigl(V^1\,h_j^0\bigr),
\end{equation}

\noindent where $V^{l}$ denotes $l$-th layer Value matrix associated with $t$;  $\alpha^{l}_{t,j}$  denotes $l$-th layer attention score that $t$ assigns to its $j$-th token in the casual attention window of size $N$; and $h_j^0$ denotes the initial hidden state of the $j$-th that $t$ attends to. As shown in \Cref{eq:h_1}, the windowed casual attention mechanism restricts $t$ to only attend to tokens associated with its preceding $n$ interactions (casual attention window of size $N$), so no hidden state leakage occurs yet.
\vspace{-0.1cm}
\begin{equation}
    \begin{aligned}
        \label{eq:h_2}
        \textbf{Layer 2:}
        \quad
        \textbf{h}_t^2
        &= \sum_{j=t-N}^{t}
             \alpha_{t,j}^2 \;
             V^2 \Bigl[
               \sum_{k=j-N}^{j}
                 \alpha_{j,k}^1 \,\bigl(V^1\,h_k^0\bigr)
             \Bigr]
        \\[2pt]
        &= \sum_{j=t-N}^{t}\;\sum_{k=j-N}^{j}
             \Bigl[\alpha_{t,j}^2\,\alpha_{j,k}^1\Bigr]\;
             \Bigl[V^2\,V^1\,h_k^0\Bigr].
    \end{aligned}    
\end{equation}

However, as shown in \Cref{eq:h_2}, starting from the second layer, the hidden states of $t$ start incorporating context interactions as far back as $t - 2N$, allowing $\textbf{h}_t^2$ to peek beyond the intended context window of size $N$. This phenomenon exacerbates as the number of forwarding layers increases:

\medskip

\textbf{ Final Layer l:}
\begin{equation}
    \label{eq:h_final}
    \textbf{h}_t^l
    =
    \sum_{j_1 = t-N}^{\,t}
    \;\dots
    \;\sum_{j_{l-1} = j_{l-2} - N}^{\,j_{l-2}}
    \Bigl[
      \alpha_{t,j_1}^l
      \,\dots
      \,\alpha_{\,j_{l-1},\,j_l}^{\,1}
    \Bigr]
    \times
    \Bigl[
      V^l\
      \,\dots
      \,V^1
      \;h_{\,j_l}^{\,0}
    \Bigr].
\end{equation}

This raises a critical discrepancy between training and inference. During inference, DTI adopts the conventional sliding-window approach to ensure inference efficiency. Each prompt only contains a total $n$ context interactions followed by an arbitrary target interaction $t$. Since there are only a total of $n$ preceding interactions for $t$ to attend, the final hidden state of $t$ will be solely based on its preceding $n$ context interactions; whereas during training the final hidden state of $t$ will encode information from interactions beyond the intended context window.

\smallskip
\noindent\textbf{Solution.}
To prevent the LLM from overly fitting on the leaked hidden states, we propose a distance-based forgetting mechanism to partially interpolate the final hidden state of each context interaction with its initial hidden states. From \Cref{eq:h_2}, it is evident that within the allowed context window ${t-N,…,t}$, interactions closer to $t$ (e.g., $t-1$) carry less leaked information compared to interactions farther from $t$ (e.g., $t-N$) which include more indirect contributions from earlier interactions beyond the intended context window. In other words, during training, hidden states of context interactions appearing early in the context would resemble less similar to their initial hidden states. However, during inference, hidden states of earlier context interactions have a much stronger resemblance to their initial hidden states, since there are simply no (or only a few) preceding interactions for them to attend to. To mitigate this discrepancy, we apply a linear interpolation to partially reset the final hidden state of a context interaction to its initial hidden state, based on the its distance from the target interaction. Formally, let $d$ be the distance from the context interaction to the target interaction, we define the linear interpolation to be: 
\begin{equation*}
    \textbf{h}_c = \alpha(d) \cdot \textbf{h}_{c}^{\text{init}} + (1 - \alpha(d)) \cdot \textbf{h}_{\text{c}},
\end{equation*}

\noindent where $\textbf{h}_c$ and $\textbf{h}_c^{init}$ respectively denotes the final and initial hidden state of an arbitrary context interaction $c$; and $\alpha(d)$ denotes the logistical distance-based interpolation ratio defined as: 
\begin{equation*}
    \alpha(d) 
    = y_{\min}
    + \Bigl(y_{\max} - y_{\min}\Bigr)
      \cdot \frac{1}{1 + \exp\bigl[-\bigl(d - N/2\bigr)\bigr]},
\end{equation*}

\noindent where $y_{min}$ and $y_{max}$ together denote the range of the logistical interpolation ratio; $d$ denotes the distance from $c$ to the target interaction $t$; and $N/2$ denotes the midpoint of the context window.

\subsection{Positional Bias Overfitting}
\smallskip
\noindent\textbf{Cause.}
Another critical issue with adopting the streaming prompt formulation for training and the sliding-window prompt formulation for inference is caused by the difference between the position ids of the [SUM] tokens in training and inference prompts. Recall that in the streaming prompt formulation, DTI appends $k$ target interactions after the first $n$ context interactions to form a prompt with $n+k$ interactions. In order to train the LLM with these streaming prompts, a [SUM] token is inserted immediately after each target interaction to be used to compute the cross-entropy loss with the corresponding label of each target interaction. However, since we are using the sliding-window approach for inference to avoid incurring additional inference overhead, each inference prompt only contains one target interaction, and thus only one [SUM] appended at the end. Due to this structural difference between training and inference, the LLM could easily overfit onto the position id information during training, causing dramatic performance degradation during inference if left unattended. 

Formally, let $\textbf{s}^{1}, ..., \textbf{s}^{k}$ denote the hidden states of all $k$ [SUM] tokens in a training prompt. For an arbitrary $\textbf{s}^i$, in modern LLMs \cite{dubey2024llama, team2024gemma, yang2024qwen2, liu2024deepseek}, its position id is generally encoded using either absolute positional embedding \cite{waswani2017attention} or Rotary Position Embedding (RoPE) \cite{su2024roformer} as follows:
\begin{equation}
    \textbf{e}_{s^{i}} =
        \begin{cases}
        \textbf{e}_{s^{i}} + \textbf{p}_{s_i}, & \text{if using absolute position embedding}, \\[3pt]
        \textbf{e}_{s} \times \textbf{R}(p_{s_i}), & \text{if using RoPE}.
        \end{cases}
\end{equation}

\noindent where $\textbf{p}_{s_i}$ denotes the positional embedding of $s_i$, and $\textbf{R}(p_{s_i})$ denotes the rotation matrix defined as follows:

\begin{equation*}
\mathbf{R}(p_{s_i}) = \begin{bmatrix} \cos(p_{s_i}\theta) & -\sin(p_{s_i}\theta) \\ \sin(p_{s_i}\theta) & \cos(p_{s_i}\theta) \end{bmatrix} \quad  \theta = c^{-\frac{1}{d_{\text{model}}}},
\end{equation*}

\noindent where $p_{s_i}$ denotes the position id of the $i$-th [SUM], $s_i$, in the training prompt. And $\theta$ denotes the scaling factor together with $c$ being a large constant and $d_{model}$ being the embedding dimensionality of the LLM. 

Regardless of the type of positional encoding used, the position id $p_{s_i}$ of each [SUM] token directly influences how its initial hidden state is altered or rotated during training. In other words, [SUM] tokens that appear in different places in the training prompt now have different representations / hidden states to the LLM. During training, the LLM can easily learn to overfit on these differences in positional encoding among different [SUM] tokens when making CTR predictions, leading to worse performance during inference.

\smallskip
\noindent\textbf{Solution.}
To address the issue of position id overfitting, DTI eliminates the use of position ids for all [SUM] tokens so that no direct positional encoding will be applied on the their initial hidden states; instead, DTI applies Attention with Linear Biases (ALiBi) \cite{press2021train} to directly fuse in the notion of position in computed attention score matrix of each [SUM] token. Formally, the attention scores are computed as follows:
\[
\mathrm{Attention}(Q, K, V)
= \mathrm{softmax}\,\Bigl(
    \frac{Q K^\top}{\sqrt{d_k}}
    + \text{Bias}(p, q)\Bigr) \, V,
\]
where and $p$ and $q$ respectively denotes the position ids of the query and the key token, and $d$ denotes the latent dimensionality. The ALiBi Bias term is defined as follows:
\begin{equation*}
    \text{Bias}(p, q) = \alpha \times (p - q),
\end{equation*}
where $\alpha$ is the ALiBi slope to be tuned. This formulation allows the LLM to treat all [SUM] tokens (target interactions) equally regardless of their positions in the prompt, achieving an alignment with inference. Meanwhile, the positional information is still preserved by adding in relative distance bias in the attention score matrices.

\begin{table*}[ht]
  \captionsetup{skip=3pt}
  \caption{CTR Prediction Performance Comparison between DTI and the traditional sliding-window (SW) training paradigm, with AUC $\uparrow$, Log Loss $\downarrow$, and F1 Score $\uparrow$ as metrics. The best result is given in bold, and the second-best value is underlined. Rel.Imp denotes the relative AUC increase or decrease rate in term of the  traditional sliding-window training paradigm. DTI$^{-}$ denotes the DTI paradigm without addressing the two bottlenecks, and DTI denotes the complete paradigm with the integration of the two proposed solutions.}
  \begin{adjustbox}
  {width=2.1\columnwidth,center}
    \begin{tabular}{l c c c c | c c c c | c c c c} 
      \hline
      \multicolumn{1}{c}{ } & \multicolumn{4}{c|}{\textbf{ML-1M}} & \multicolumn{4}{c|}{\textbf{Amazon CDs}} & \multicolumn{4}{c}{\textbf{Amazon Kindles}} \\
      \multicolumn{1}{c}{ } & AUC & Log Loss & F1 & Rel. Imp & AUC & Log Loss & F1 & Rel. Imp & AUC & Log Loss & F1 & Rel. Imp \\
      \hline
      SW (K=1) & \underline{0.8110} & \textbf{0.5242} & 0.7657 & \textbf{---}  & 0.8596 & 0.4025 & 0.8717 & \textbf{---} & 0.8689 & 0.4495 & 0.8106 & \textbf{---} \\
      \hline\hline
      DTI$^{-}$ (k=5) & 0.8103 & 0.5268 & 0.7641 & -0.09\% & 0.8543 & 0.4062 & 0.8779 & -0.62\% & 0.8685 & 0.4528 & 0.8044 & -0.05\% \\
      DTI$^{-}$ (k=10) & 0.8090 & 0.5447 & 0.7327 & -0.25\% & 0.8525 & 0.4054 & 0.8753 & -0.83\% & 0.8670 & 0.4648 & 0.7869 & -0.22\%  \\
      DTI$^{-}$ (k=20) & 0.8071 & 0.5410 & 0.7290 & -0.48\% & 0.8485 & 0.4125 & 0.8749 & -1.29\% & 0.8669 & 0.4599 & 0.7816 & -0.23\%  \\
      DTI$^{-}$ (k=30) & 0.8048 & 0.5562 & 0.7273 & -0.76\% & 0.8490 & 0.4137 & 0.8775 & -1.23\% & 0.8670 & 0.4703 & 0.7832 & -0.22\%  \\
      DTI$^{-}$ (k=40) & 0.8051 & 0.5526 & 0.7280 & -0.73\% & 0.8452 & 0.4149 & 0.8737 & -1.68\% & 0.8659 & 0.4708 & 0.7859 & -0.35\%  \\
      DTI$^{-}$ (k=50) & 0.8043 & 0.5506 & 0.7042 & -0.83\% & 0.8394 & 0.4213 & 0.8707 & -2.35\% & 0.8654 & 0.4721 & 0.7827 & -0.40\%  \\
      \hline\hline
      DTI (k=5) & \textbf{0.8118} & \underline{0.5255} & 0.7683 & +0.10\% & 0.8606 & 0.3993 & \underline{0.8792} & +0.17\% & \textbf{0.8702} & \textbf{0.4472} & 0.8105 & +0.15\% \\
      DTI (k=10) & 0.8101 & 0.5264 & \underline{0.7687} & -0.11\% & \textbf{0.8615} & \textbf{0.3983} & 0.8749 & +0.22\% & 0.8698 & \underline{0.4489} & 0.8086 & +0.10\% \\
      DTI (k=20) & 0.8102 & 0.5257 & 0.7642 & -0.10\% & 0.8599 & \underline{0.3988} & 0.8767 & +0.03\% & 0.8681 & 0.4519 & 0.8058 & -0.09\% \\ 
      DTI (k=30) & 0.8103 & 0.5265 & 0.7667 & -0.09\% & 0.8581 & 0.4014 & 0.8758 & -0.17\% & 0.8694 & 0.4500 & \underline{0.8107} & +0.06\% \\ 
      DTI (k=40) & 0.8099 & 0.5285 & 0.7610 & -0.14\% & \underline{0.8608} & 0.4003 & \textbf{0.8810} & +0.14\% & \underline{0.8699} & 0.4495 & \textbf{0.8113} & +0.12\% \\
      DTI (k=50) & 0.8104 & 0.5251 & \textbf{0.7689} & -0.07\% & 0.8601 & 0.3984 & 0.8757 & +0.06\% & 0.8692 & 0.4506 & 0.8047 & +0.03\% \\ 
      \hline
    \end{tabular}
  \end{adjustbox}
  \label{tab:DTI_performance}
\end{table*}

\section{EXPERIMENTS}
To validate the effectiveness of DTI, we conduct experiments on three widely adopted datasets to answer three key research questions: \textbf{RQ1.} Does the proposed training paradigm effectively scale up to large values of $k$ with the integration of solutions for the two identified bottlenecks? \textbf{RQ2.} How well does the predicted training FLOPs reduction in \Cref{eq:flop_red} translate into reduction in training time until convergence when empirically tested? \textbf{RQ3.} How does the solution of each of the two bottlenecks contribute to maintaining the LLM's CTR performance as $k$ scales up?

\subsection{Experiment Setup}
\begin{table}
 .\begin{center}
  \setlength{\tabcolsep}{5pt}
    \captionsetup{skip=5pt}
    \caption{Dataset statistics}
    \label{tab:table1}
    \setlength{\tabcolsep}{10pt}
    \begin{tabular}{l p{1cm} p{1cm} p{1cm} p{1cm} r}
      \hline
      Dataset & \#User & \#Item & \#Sample & Sparsity \\
      \hline  
      ML-1M & 6,040 & 3,706 & 1,000,209 & 0.9553 \\
      Kindles & 14,051 & 98,772 & 1,026,351 & 0.9993 \\
      CDs & 6,741 & 73,403 & 492,333 & 0.9990 \\
      \hline
    \end{tabular}
    \end{center}
\end{table}

\noindent\textbf{Dataset.} We conduct experiments on three widely adopted data sets: MovieLens 1M \cite{harper2015movielens}, Amazon-Kindles \cite{ni2019justifying}, and Amazon-CDs \cite{ni2019justifying} For all three data sets, we follow the common recommendation evaluation setting \cite{liu2024mamba4rec} to prune items with fewer than 5 interactions and split user-item interactions in a ratio of 8: 1: 1 according to the timestamp for training, validation, and testing, to mimic the sequential (user-behavior-based) CTR prediction setting. The statistics of the three data sets are shown in \Cref{tab:table1}. 

\noindent\textbf{Evaluation Metrics.} To validate the effectiveness of DTI, we adopt the AUC (area under the ROC curve), Log Loss, and the F1 score as evaluation metrics. As indicated in previous studies \cite{lin2024rella, lian2018xdeepfm, wang2021dcn}, a slightly higher AUC or lower Log Loss (e.g., 0.001) can be considered as a significant improvement in CTR prediction performance.

\subsection{DTI Training Performance (RQ1)}
To answer \textbf{RQ1}, we compare the CTR prediction performance of LLMs trained using the traditional sliding-window paradigm and the LLMs trained using the proposed DTI. As shown in \Cref{tab:DTI_performance}, without addressing the identified bottlenecks, the performance of the trained LLM consistently decreases as the value of $k$ (number of target interactions per streaming prompt) increases. This is because a larger value of $k$ creates a stronger discrepancy between training and inference. Interestingly, compared to the decrease in AUC, a sharper degradation in both Log Loss and F1 Score is observed as the value of $k$ increases. This observation suggests that while two identified bottlenecks -- hidden-states leakage and positional overfitting -- heavily affects the predicts logits of the labels, their relative rank still holds to a certain extend -- further showing the robustness of LLMs on ranking-based recommendations. With the integration of the two proposed solutions, one for addressing each bottleneck, DTI incurs slight performance degradation on MoveLens-1M dataset, but achieves a slightly better performance on both the Amazon CDs and Kindles dataset. Such observations are rather encouraging as they suggest: a). DTI holds its main objective of reducing the training time without sacrificing performance; and b). the proposed solutions for the two bottlenecks are  effective. In RQ2 and RQ3, we further illustrate details on the reduced training time until convergence and perform ablations to see how each of the two proposed bottleneck solution contributes to maintaining DTI's performance when a large value of $k$ is chosen.

\subsection{Training Time Reduction (RQ2)}
\begin{table*}[ht]
  \caption{Training Time Comparison between DTI and the widely adopted sliding-window (SW) training paradigm, with the ``wall clock time $\downarrow$'' as metric. The best result is given in bold. Rel.Red denotes the relative training time reduction percentage. Models are trained for 10 epochs with the early stopping strategy.}
  \begin{adjustbox}
  {width=2.1\columnwidth,center}
    \begin{tabular}{l c c c c| c c c c| c c c c} 
      \hline
      \multicolumn{1}{c}{ } & \multicolumn{4}{c|}{\textbf{ML-1M}} & \multicolumn{4}{c|}{\textbf{CDs}} & \multicolumn{4}{c}{\textbf{ Kindles}} \\
      \multicolumn{1}{c}{ } & AUC & Log Loss & Time & Rel. Red & AUC & Log Loss & Time & Rel. Red & AUC & Log Loss & Time & Rel. Red \\
      \hline
      SW (K=1)  & \textbf{0.8110} & \textbf{0.5242} & 69.56h & \textbf{---} & 0.8596 & 0.4025 & 37.84h & \textbf{---} & 0.8689 & 0.4495 & 70.50h & \textbf{---}\\
      \hline
      DTI (k=10) & 0.8101 & 0.5264 & 11.13h & $\downarrow84.00\%$ & \textbf{0.8615} & \textbf{0.3983} & 07.56h & $\downarrow80.02\%$ & 0.8698 & \textbf{0.4489} & 11.98h & $\downarrow83.00\%$ \\
      DTI (k=30) & 0.8103 & 0.5265 & 06.63h & $\downarrow90.47\%$ & 0.8581 & 0.4014 & 04.75h & $\downarrow87.45\%$ & \textbf{0.8694} & 0.4500 & 06.20h &  $\downarrow91.21\%$ \\ 
      DTI (k=50) & 0.8104 & 0.5251 & \textbf{05.37h} & $\downarrow\textbf{92.28\%}$ & 0.8601 & 0.3984 & \textbf{03.42h} & $\downarrow\textbf{90.96\%}$ & 0.8692 & 0.4506 & \textbf{05.31h} & $\downarrow\textbf{92.47\%}$ \\ 
      \hline
    \end{tabular}
  \end{adjustbox}
  \label{tab:DTI_time_red}
\end{table*}

In \Cref{eq:flop_red}, we quantify the training FLOPs reduction in terms of the length of a sliding-window prompt (N), the length a streaming prompt (K+N), and most importantly the number of target interactions ($k$) included in one streaming prompt. To test how the predicted FLOPs reduction translate into training time reduction, we empirically compare the training time until converge between the DTI and the traditional sliding-window training paradigm. As shown in \Cref{tab:DTI_time_red}, a consistent decrease in training time is observed across all three datasets as the value of $K$ increases. More interestingly, the measured training time reduction aligns well with the calculated FLOPs reduction especially on larger datasets. For instance, when plugging in $50c$ for $K$ and $20c$ for $N$, where $c$ is a constant that represents the number of tokens needed  to represent an interaction, \Cref{eq:flop_red} gets a FLOPs reduction of $14.28$ times which translates surprisingly well into observed training time reductions across all three datasets, especially for MovieLens-1M and Amazon Kindles. It is also important to note that such significant training time reduction does not incur any any performance sacrifice or inference inefficiency.  

\subsection{Ablation Studies (RQ3)}
\begin{figure}
  \includegraphics[width=.5\textwidth]{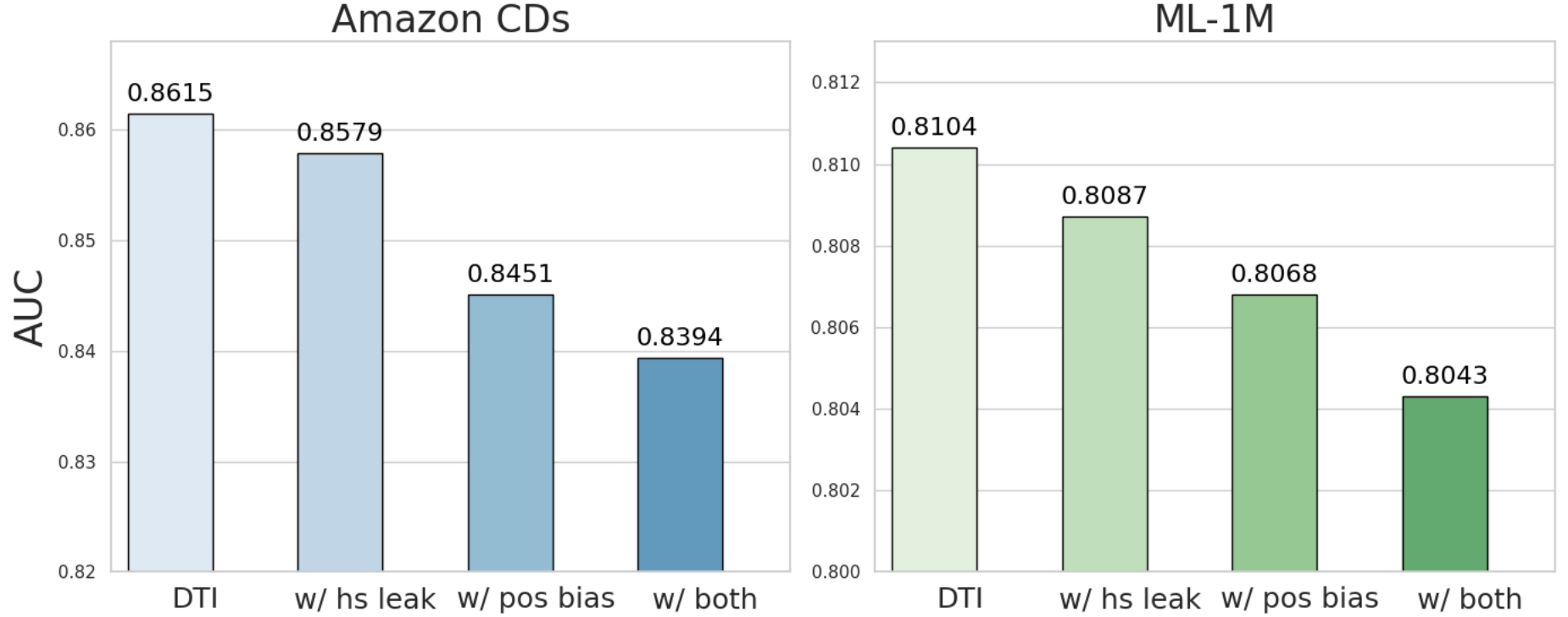}
  \caption{Ablation Studies to analyze the effectiveness of each of the two solutions for the two bottlenecks. {w/ hs leak} denotes the version that only addresses positional bias overfitting. {w/pos bias} denotes the version that only addresses hidden state leakages. {w/both} denotes addressing neither.}
  \label{fig:DTI_ablation}
\end{figure}

We conduct ablation studies on the Amazon CDs dataset and the MovieLens-1M dataset to respectively analyze the effectiveness of each of the proposed solutions for the two identified bottlenecks - namely, hidden state leakage and positional bias overfitting. As shown in \Cref{fig:DTI_ablation}, DTI is much more prone to positional bias overfitting than hidden state leakage, as addressing the positional bias overfitting alone results in significant improvement in AUC. Nevertheless, both issues affect the performance of the trained LLM so should both be addressed to ensure optimal training results. 

\section{RELATED WORKS}
In general, the recent application of LLMs in Recommendation systems can be classified into two main directions -- feature augmentation \cite{christakopoulou2023large, li2023taggpt, mysore2023large, liu2023first} and recommendation generation \cite{bao2023tallrec, yang2023palr, hou2024large}. For feature augmentation, LLMs are typically used to generate additional supplementary attributes based on text-based user profiles and item descriptions. For example, KAR \cite{xi2024towards} uses a novel factorization prompting technique to generate additional features to enhance the data set and promote recommendation performance in a model-agnostic way. Moreover, LLMs are also adopted as auxiliary textual feature encoders to enrich the user/item representations with semantic information. For instance, U-BERT \cite{qu2016product} enhances the user representation by encoding review texts into dense vectors via BERT. RLMRec \cite{ren2024representation} utilizes contrastive and generative alignment techniques to align CF-side relational embedding with LLM-side semantics, enhancing the quality of user/item representations. Compared to feature augmentation, the application of LLMs directly for recommendation generation has received more recent attention. Many recent studies \cite{hua2023index, kang2023llms, li2023text} have explored the use of LLMs' in-context learning abilities for recommendation tasks \cite{dong2022survey}. For example, Hou et al. \cite{hou2024large} and Gao et al \cite{gao2023chat} showcase LLMs' impressive zero-shot ranking capabilities. Interestingly, due to the significant gap between the pre-training data and the recommendation task, these approaches often underperform traditional recommendation models \cite{wei2023larger}. To bridge this gap, many studies \cite{bao2023tallrec, zhang2023recommendation, feng2023large} attempt to finetune the LLM to better adapt to the recommendation domain with textualized user-item interaction data. However, as noted in \cite{lin2024rella, collm}, these approaches still underperform well-tuned conventional recommendation models due to the difficulty for LLMs in capturing the highly crucial underlying collaborative filtering (CF) signals. To address this shortcoming, many CF fusion strategies \cite{bao2023bi, zhu2024collaborative, zheng2024adapting} have recently been proposed to enhance the fine-tuning process of LLMs. For instance, \cite{bao2023bi} ensembles LLMs with collaborative models to prevent LLMs from relying solely on semantic priors and overlooking collaborative information. However, as pointed out by \cite{collm}, this two-stage integration strategy limits the full potential of LLMs. To address this, \cite{collm} proposes learning a linear mapping function to project the learned ID embeddings onto the LLM space, causing a more seemingless integration of collaborative signals. However, the effectiveness of such an approach is heavily based on the learned mapping function, due to the huge distribution gap between the data used to pretrain the LLM and the user-item interaction data used to learn the ID embeddings. In addition, \cite{zhu2024collaborative} proposes to expand the original LLM vocab by adding a unique embedding for each user and item. 

One work that inspires the streaming-prompt formulation is the HSTU \cite{zhai2024actions} framework proposed by Zhai et al. Specifically, HSTU \cite{zhai2024actions} proposes the M-FALCON algorithm to enhance the inference efficiency via masked attention and novel caching techniques. Inspired by it, we propose the streaming prompt formulation to parallelize the training of multiple targets. More importantly, our work identifies two main bottlenecks - hidden state leakage and positional bias overfitting - that lead to CTR performance degradation as more targets are being included in one training prompt. To address this issue, we propose tailored solutions for both and empirically show that the integration of the proposed solutions allows DTI to parallelize the training of up to 50 targets without sacrificing performance. 

\section{CONCLUSION}
In this work, we propose a novel \textbf{D}ynamic \textbf{T}arget \textbf{I}solation (DTI) paradigm that structurally parallelizes the training of $k$ (where $k >> 1$) target interactions in one single streaming prompt. Compare to the conventional sliding-window training paradigm that construct a distinct prompt for each target interaction, DTI reduces the total number of training prompts by a factor of $k$. To counter the additional computational and memory overhead brought by quadratic scaling due to the increased prompt length, DTI adopts a windowed casual attention mechanism to restrict each target interaction to only attend to tokens associated with its previous $n$ interactions. To minimize the discrepancy between training and inference, namely hidden-state leakage and positional bias overfitting, between training and inference, DTI further integrates a distance-based hidden state forgetting mechanism along with relative-position ids to ensure training effectiveness without compromising inference efficiency. Together, we theoretically and theoretically show that DTI dramatically reduces training FLOPs and training times needed until convergence, while incurring minimal to none compromise in CTR prediction performance, when tested on three widely adopted public datasets. 

\bibliographystyle{ACM-Reference-Format}
\bibliography{ref}


\begin{thebibliography}{51}


\ifx \showCODEN    \undefined \def \showCODEN     #1{\unskip}     \fi
\ifx \showDOI      \undefined \def \showDOI       #1{#1}\fi
\ifx \showISBNx    \undefined \def \showISBNx     #1{\unskip}     \fi
\ifx \showISBNxiii \undefined \def \showISBNxiii  #1{\unskip}     \fi
\ifx \showISSN     \undefined \def \showISSN      #1{\unskip}     \fi
\ifx \showLCCN     \undefined \def \showLCCN      #1{\unskip}     \fi
\ifx \shownote     \undefined \def \shownote      #1{#1}          \fi
\ifx \showarticletitle \undefined \def \showarticletitle #1{#1}   \fi
\ifx \showURL      \undefined \def \showURL       {\relax}        \fi
\providecommand\bibfield[2]{#2}
\providecommand\bibinfo[2]{#2}
\providecommand\natexlab[1]{#1}
\providecommand\showeprint[2][]{arXiv:#2}

\bibitem[Bao et~al\mbox{.}(2023a)]%
        {bao2023bi}
\bibfield{author}{\bibinfo{person}{Keqin Bao}, \bibinfo{person}{Jizhi Zhang}, \bibinfo{person}{Wenjie Wang}, \bibinfo{person}{Yang Zhang}, \bibinfo{person}{Zhengyi Yang}, \bibinfo{person}{Yancheng Luo}, \bibinfo{person}{Chong Chen}, \bibinfo{person}{Fuli Feng}, {and} \bibinfo{person}{Qi Tian}.} \bibinfo{year}{2023}\natexlab{a}.
\newblock \showarticletitle{A bi-step grounding paradigm for large language models in recommendation systems}.
\newblock \bibinfo{journal}{\emph{arXiv preprint arXiv:2308.08434}} (\bibinfo{year}{2023}).
\newblock


\bibitem[Bao et~al\mbox{.}(2023b)]%
        {bao2023tallrec}
\bibfield{author}{\bibinfo{person}{Keqin Bao}, \bibinfo{person}{Jizhi Zhang}, \bibinfo{person}{Yang Zhang}, \bibinfo{person}{Wenjie Wang}, \bibinfo{person}{Fuli Feng}, {and} \bibinfo{person}{Xiangnan He}.} \bibinfo{year}{2023}\natexlab{b}.
\newblock \showarticletitle{Tallrec: An effective and efficient tuning framework to align large language model with recommendation}. In \bibinfo{booktitle}{\emph{Proceedings of the 17th ACM Conference on Recommender Systems}}. \bibinfo{pages}{1007--1014}.
\newblock


\bibitem[Christakopoulou et~al\mbox{.}(2023)]%
        {christakopoulou2023large}
\bibfield{author}{\bibinfo{person}{Konstantina Christakopoulou}, \bibinfo{person}{Alberto Lalama}, \bibinfo{person}{Cj Adams}, \bibinfo{person}{Iris Qu}, \bibinfo{person}{Yifat Amir}, \bibinfo{person}{Samer Chucri}, \bibinfo{person}{Pierce Vollucci}, \bibinfo{person}{Fabio Soldo}, \bibinfo{person}{Dina Bseiso}, \bibinfo{person}{Sarah Scodel}, {et~al\mbox{.}}} \bibinfo{year}{2023}\natexlab{}.
\newblock \showarticletitle{Large language models for user interest journeys}.
\newblock \bibinfo{journal}{\emph{arXiv preprint arXiv:2305.15498}} (\bibinfo{year}{2023}).
\newblock


\bibitem[Dong et~al\mbox{.}(2022)]%
        {dong2022survey}
\bibfield{author}{\bibinfo{person}{Qingxiu Dong}, \bibinfo{person}{Lei Li}, \bibinfo{person}{Damai Dai}, \bibinfo{person}{Ce Zheng}, \bibinfo{person}{Jingyuan Ma}, \bibinfo{person}{Rui Li}, \bibinfo{person}{Heming Xia}, \bibinfo{person}{Jingjing Xu}, \bibinfo{person}{Zhiyong Wu}, \bibinfo{person}{Tianyu Liu}, {et~al\mbox{.}}} \bibinfo{year}{2022}\natexlab{}.
\newblock \showarticletitle{A survey on in-context learning}.
\newblock \bibinfo{journal}{\emph{arXiv preprint arXiv:2301.00234}} (\bibinfo{year}{2022}).
\newblock


\bibitem[Dubey et~al\mbox{.}(2024)]%
        {dubey2024llama}
\bibfield{author}{\bibinfo{person}{Abhimanyu Dubey}, \bibinfo{person}{Abhinav Jauhri}, \bibinfo{person}{Abhinav Pandey}, \bibinfo{person}{Abhishek Kadian}, \bibinfo{person}{Ahmad Al-Dahle}, \bibinfo{person}{Aiesha Letman}, \bibinfo{person}{Akhil Mathur}, \bibinfo{person}{Alan Schelten}, \bibinfo{person}{Amy Yang}, \bibinfo{person}{Angela Fan}, {et~al\mbox{.}}} \bibinfo{year}{2024}\natexlab{}.
\newblock \showarticletitle{The llama 3 herd of models}.
\newblock \bibinfo{journal}{\emph{arXiv preprint arXiv:2407.21783}} (\bibinfo{year}{2024}).
\newblock


\bibitem[Feng et~al\mbox{.}(2023)]%
        {feng2023large}
\bibfield{author}{\bibinfo{person}{Yue Feng}, \bibinfo{person}{Shuchang Liu}, \bibinfo{person}{Zhenghai Xue}, \bibinfo{person}{Qingpeng Cai}, \bibinfo{person}{Lantao Hu}, \bibinfo{person}{Peng Jiang}, \bibinfo{person}{Kun Gai}, {and} \bibinfo{person}{Fei Sun}.} \bibinfo{year}{2023}\natexlab{}.
\newblock \showarticletitle{A large language model enhanced conversational recommender system}.
\newblock \bibinfo{journal}{\emph{arXiv preprint arXiv:2308.06212}} (\bibinfo{year}{2023}).
\newblock


\bibitem[Gao et~al\mbox{.}(2023)]%
        {gao2023chat}
\bibfield{author}{\bibinfo{person}{Yunfan Gao}, \bibinfo{person}{Tao Sheng}, \bibinfo{person}{Youlin Xiang}, \bibinfo{person}{Yun Xiong}, \bibinfo{person}{Haofen Wang}, {and} \bibinfo{person}{Jiawei Zhang}.} \bibinfo{year}{2023}\natexlab{}.
\newblock \showarticletitle{Chat-rec: Towards interactive and explainable llms-augmented recommender system}.
\newblock \bibinfo{journal}{\emph{arXiv preprint arXiv:2303.14524}} (\bibinfo{year}{2023}).
\newblock


\bibitem[Geng et~al\mbox{.}(2022)]%
        {geng2022recommendation}
\bibfield{author}{\bibinfo{person}{Shijie Geng}, \bibinfo{person}{Shuchang Liu}, \bibinfo{person}{Zuohui Fu}, \bibinfo{person}{Yingqiang Ge}, {and} \bibinfo{person}{Yongfeng Zhang}.} \bibinfo{year}{2022}\natexlab{}.
\newblock \showarticletitle{Recommendation as language processing (rlp): A unified pretrain, personalized prompt \& predict paradigm (p5)}. In \bibinfo{booktitle}{\emph{Proceedings of the 16th ACM Conference on Recommender Systems}}. \bibinfo{pages}{299--315}.
\newblock


\bibitem[Harper and Konstan(2015)]%
        {harper2015movielens}
\bibfield{author}{\bibinfo{person}{F~Maxwell Harper} {and} \bibinfo{person}{Joseph~A Konstan}.} \bibinfo{year}{2015}\natexlab{}.
\newblock \showarticletitle{The movielens datasets: History and context}.
\newblock \bibinfo{journal}{\emph{Acm transactions on interactive intelligent systems (tiis)}} \bibinfo{volume}{5}, \bibinfo{number}{4} (\bibinfo{year}{2015}), \bibinfo{pages}{1--19}.
\newblock


\bibitem[Hou et~al\mbox{.}(2024)]%
        {hou2024large}
\bibfield{author}{\bibinfo{person}{Yupeng Hou}, \bibinfo{person}{Junjie Zhang}, \bibinfo{person}{Zihan Lin}, \bibinfo{person}{Hongyu Lu}, \bibinfo{person}{Ruobing Xie}, \bibinfo{person}{Julian McAuley}, {and} \bibinfo{person}{Wayne~Xin Zhao}.} \bibinfo{year}{2024}\natexlab{}.
\newblock \showarticletitle{Large language models are zero-shot rankers for recommender systems}. In \bibinfo{booktitle}{\emph{European Conference on Information Retrieval}}. Springer, \bibinfo{pages}{364--381}.
\newblock


\bibitem[Hu et~al\mbox{.}(2021)]%
        {hu2021lora}
\bibfield{author}{\bibinfo{person}{Edward~J Hu}, \bibinfo{person}{Yelong Shen}, \bibinfo{person}{Phillip Wallis}, \bibinfo{person}{Zeyuan Allen-Zhu}, \bibinfo{person}{Yuanzhi Li}, \bibinfo{person}{Shean Wang}, \bibinfo{person}{Lu Wang}, {and} \bibinfo{person}{Weizhu Chen}.} \bibinfo{year}{2021}\natexlab{}.
\newblock \showarticletitle{Lora: Low-rank adaptation of large language models}.
\newblock \bibinfo{journal}{\emph{arXiv preprint arXiv:2106.09685}} (\bibinfo{year}{2021}).
\newblock


\bibitem[Hua et~al\mbox{.}(2023)]%
        {hua2023index}
\bibfield{author}{\bibinfo{person}{Wenyue Hua}, \bibinfo{person}{Shuyuan Xu}, \bibinfo{person}{Yingqiang Ge}, {and} \bibinfo{person}{Yongfeng Zhang}.} \bibinfo{year}{2023}\natexlab{}.
\newblock \showarticletitle{How to index item ids for recommendation foundation models}. In \bibinfo{booktitle}{\emph{Proceedings of the Annual International ACM SIGIR Conference on Research and Development in Information Retrieval in the Asia Pacific Region}}. \bibinfo{pages}{195--204}.
\newblock


\bibitem[Kang and McAuley(2018)]%
        {kang2018self}
\bibfield{author}{\bibinfo{person}{Wang-Cheng Kang} {and} \bibinfo{person}{Julian McAuley}.} \bibinfo{year}{2018}\natexlab{}.
\newblock \showarticletitle{Self-attentive sequential recommendation}. In \bibinfo{booktitle}{\emph{2018 IEEE international conference on data mining (ICDM)}}. IEEE, \bibinfo{pages}{197--206}.
\newblock


\bibitem[Kang et~al\mbox{.}(2023)]%
        {kang2023llms}
\bibfield{author}{\bibinfo{person}{Wang-Cheng Kang}, \bibinfo{person}{Jianmo Ni}, \bibinfo{person}{Nikhil Mehta}, \bibinfo{person}{Maheswaran Sathiamoorthy}, \bibinfo{person}{Lichan Hong}, \bibinfo{person}{Ed Chi}, {and} \bibinfo{person}{Derek~Zhiyuan Cheng}.} \bibinfo{year}{2023}\natexlab{}.
\newblock \showarticletitle{Do llms understand user preferences? evaluating llms on user rating prediction}.
\newblock \bibinfo{journal}{\emph{arXiv preprint arXiv:2305.06474}} (\bibinfo{year}{2023}).
\newblock


\bibitem[Kim et~al\mbox{.}(2024)]%
        {kim2024large}
\bibfield{author}{\bibinfo{person}{Sein Kim}, \bibinfo{person}{Hongseok Kang}, \bibinfo{person}{Seungyoon Choi}, \bibinfo{person}{Donghyun Kim}, \bibinfo{person}{Minchul Yang}, {and} \bibinfo{person}{Chanyoung Park}.} \bibinfo{year}{2024}\natexlab{}.
\newblock \showarticletitle{Large language models meet collaborative filtering: An efficient all-round llm-based recommender system}. In \bibinfo{booktitle}{\emph{Proceedings of the 30th ACM SIGKDD Conference on Knowledge Discovery and Data Mining}}. \bibinfo{pages}{1395--1406}.
\newblock


\bibitem[Kingma(2014)]%
        {kingma2014adam}
\bibfield{author}{\bibinfo{person}{Diederik~P Kingma}.} \bibinfo{year}{2014}\natexlab{}.
\newblock \showarticletitle{Adam: A method for stochastic optimization}.
\newblock \bibinfo{journal}{\emph{arXiv preprint arXiv:1412.6980}} (\bibinfo{year}{2014}).
\newblock


\bibitem[Li et~al\mbox{.}(2023a)]%
        {li2023taggpt}
\bibfield{author}{\bibinfo{person}{Chen Li}, \bibinfo{person}{Yixiao Ge}, \bibinfo{person}{Jiayong Mao}, \bibinfo{person}{Dian Li}, {and} \bibinfo{person}{Ying Shan}.} \bibinfo{year}{2023}\natexlab{a}.
\newblock \showarticletitle{Taggpt: Large language models are zero-shot multimodal taggers}.
\newblock \bibinfo{journal}{\emph{arXiv preprint arXiv:2304.03022}} (\bibinfo{year}{2023}).
\newblock


\bibitem[Li et~al\mbox{.}(2023b)]%
        {li2023text}
\bibfield{author}{\bibinfo{person}{Jiacheng Li}, \bibinfo{person}{Ming Wang}, \bibinfo{person}{Jin Li}, \bibinfo{person}{Jinmiao Fu}, \bibinfo{person}{Xin Shen}, \bibinfo{person}{Jingbo Shang}, {and} \bibinfo{person}{Julian McAuley}.} \bibinfo{year}{2023}\natexlab{b}.
\newblock \showarticletitle{Text is all you need: Learning language representations for sequential recommendation}. In \bibinfo{booktitle}{\emph{Proceedings of the 29th ACM SIGKDD Conference on Knowledge Discovery and Data Mining}}. \bibinfo{pages}{1258--1267}.
\newblock


\bibitem[Lian et~al\mbox{.}(2018)]%
        {lian2018xdeepfm}
\bibfield{author}{\bibinfo{person}{Jianxun Lian}, \bibinfo{person}{Xiaohuan Zhou}, \bibinfo{person}{Fuzheng Zhang}, \bibinfo{person}{Zhongxia Chen}, \bibinfo{person}{Xing Xie}, {and} \bibinfo{person}{Guangzhong Sun}.} \bibinfo{year}{2018}\natexlab{}.
\newblock \showarticletitle{xdeepfm: Combining explicit and implicit feature interactions for recommender systems}. In \bibinfo{booktitle}{\emph{Proceedings of the 24th ACM SIGKDD international conference on knowledge discovery \& data mining}}. \bibinfo{pages}{1754--1763}.
\newblock


\bibitem[Liao et~al\mbox{.}(2023)]%
        {liao2023llara}
\bibfield{author}{\bibinfo{person}{Jiayi Liao}, \bibinfo{person}{Sihang Li}, \bibinfo{person}{Zhengyi Yang}, \bibinfo{person}{Jiancan Wu}, \bibinfo{person}{Yancheng Yuan}, \bibinfo{person}{Xiang Wang}, {and} \bibinfo{person}{Xiangnan He}.} \bibinfo{year}{2023}\natexlab{}.
\newblock \showarticletitle{Llara: Aligning large language models with sequential recommenders}.
\newblock \bibinfo{journal}{\emph{arXiv preprint arXiv:2312.02445}} (\bibinfo{year}{2023}).
\newblock


\bibitem[Lin et~al\mbox{.}(2023)]%
        {lin2023can}
\bibfield{author}{\bibinfo{person}{Jianghao Lin}, \bibinfo{person}{Xinyi Dai}, \bibinfo{person}{Yunjia Xi}, \bibinfo{person}{Weiwen Liu}, \bibinfo{person}{Bo Chen}, \bibinfo{person}{Hao Zhang}, \bibinfo{person}{Yong Liu}, \bibinfo{person}{Chuhan Wu}, \bibinfo{person}{Xiangyang Li}, \bibinfo{person}{Chenxu Zhu}, {et~al\mbox{.}}} \bibinfo{year}{2023}\natexlab{}.
\newblock \showarticletitle{How can recommender systems benefit from large language models: A survey}.
\newblock \bibinfo{journal}{\emph{arXiv preprint arXiv:2306.05817}} (\bibinfo{year}{2023}).
\newblock


\bibitem[Lin et~al\mbox{.}(2024)]%
        {lin2024rella}
\bibfield{author}{\bibinfo{person}{Jianghao Lin}, \bibinfo{person}{Rong Shan}, \bibinfo{person}{Chenxu Zhu}, \bibinfo{person}{Kounianhua Du}, \bibinfo{person}{Bo Chen}, \bibinfo{person}{Shigang Quan}, \bibinfo{person}{Ruiming Tang}, \bibinfo{person}{Yong Yu}, {and} \bibinfo{person}{Weinan Zhang}.} \bibinfo{year}{2024}\natexlab{}.
\newblock \showarticletitle{Rella: Retrieval-enhanced large language models for lifelong sequential behavior comprehension in recommendation}. In \bibinfo{booktitle}{\emph{Proceedings of the ACM on Web Conference 2024}}. \bibinfo{pages}{3497--3508}.
\newblock


\bibitem[Liu et~al\mbox{.}(2024a)]%
        {liu2024deepseek}
\bibfield{author}{\bibinfo{person}{Aixin Liu}, \bibinfo{person}{Bei Feng}, \bibinfo{person}{Bing Xue}, \bibinfo{person}{Bingxuan Wang}, \bibinfo{person}{Bochao Wu}, \bibinfo{person}{Chengda Lu}, \bibinfo{person}{Chenggang Zhao}, \bibinfo{person}{Chengqi Deng}, \bibinfo{person}{Chenyu Zhang}, \bibinfo{person}{Chong Ruan}, {et~al\mbox{.}}} \bibinfo{year}{2024}\natexlab{a}.
\newblock \showarticletitle{Deepseek-v3 technical report}.
\newblock \bibinfo{journal}{\emph{arXiv preprint arXiv:2412.19437}} (\bibinfo{year}{2024}).
\newblock


\bibitem[Liu et~al\mbox{.}(2024b)]%
        {liu2024mamba4rec}
\bibfield{author}{\bibinfo{person}{Chengkai Liu}, \bibinfo{person}{Jianghao Lin}, \bibinfo{person}{Jianling Wang}, \bibinfo{person}{Hanzhou Liu}, {and} \bibinfo{person}{James Caverlee}.} \bibinfo{year}{2024}\natexlab{b}.
\newblock \showarticletitle{Mamba4rec: Towards efficient sequential recommendation with selective state space models}.
\newblock \bibinfo{journal}{\emph{arXiv preprint arXiv:2403.03900}} (\bibinfo{year}{2024}).
\newblock


\bibitem[Liu et~al\mbox{.}(2023)]%
        {liu2023first}
\bibfield{author}{\bibinfo{person}{Qijiong Liu}, \bibinfo{person}{Nuo Chen}, \bibinfo{person}{Tetsuya Sakai}, {and} \bibinfo{person}{Xiao-Ming Wu}.} \bibinfo{year}{2023}\natexlab{}.
\newblock \showarticletitle{A first look at llm-powered generative news recommendation}.
\newblock \bibinfo{journal}{\emph{CoRR}} (\bibinfo{year}{2023}).
\newblock


\bibitem[Minaee et~al\mbox{.}(2024)]%
        {minaee2024large}
\bibfield{author}{\bibinfo{person}{Shervin Minaee}, \bibinfo{person}{Tomas Mikolov}, \bibinfo{person}{Narjes Nikzad}, \bibinfo{person}{Meysam Chenaghlu}, \bibinfo{person}{Richard Socher}, \bibinfo{person}{Xavier Amatriain}, {and} \bibinfo{person}{Jianfeng Gao}.} \bibinfo{year}{2024}\natexlab{}.
\newblock \showarticletitle{Large language models: A survey}.
\newblock \bibinfo{journal}{\emph{arXiv preprint arXiv:2402.06196}} (\bibinfo{year}{2024}).
\newblock


\bibitem[Mysore et~al\mbox{.}(2023)]%
        {mysore2023large}
\bibfield{author}{\bibinfo{person}{Sheshera Mysore}, \bibinfo{person}{Andrew McCallum}, {and} \bibinfo{person}{Hamed Zamani}.} \bibinfo{year}{2023}\natexlab{}.
\newblock \showarticletitle{Large language model augmented narrative driven recommendations}. In \bibinfo{booktitle}{\emph{Proceedings of the 17th ACM Conference on Recommender Systems}}. \bibinfo{pages}{777--783}.
\newblock


\bibitem[Ni et~al\mbox{.}(2019)]%
        {ni2019justifying}
\bibfield{author}{\bibinfo{person}{Jianmo Ni}, \bibinfo{person}{Jiacheng Li}, {and} \bibinfo{person}{Julian McAuley}.} \bibinfo{year}{2019}\natexlab{}.
\newblock \showarticletitle{Justifying recommendations using distantly-labeled reviews and fine-grained aspects}. In \bibinfo{booktitle}{\emph{Proceedings of the 2019 conference on empirical methods in natural language processing and the 9th international joint conference on natural language processing (EMNLP-IJCNLP)}}. \bibinfo{pages}{188--197}.
\newblock


\bibitem[Pi et~al\mbox{.}(2020)]%
        {pi2020search}
\bibfield{author}{\bibinfo{person}{Qi Pi}, \bibinfo{person}{Guorui Zhou}, \bibinfo{person}{Yujing Zhang}, \bibinfo{person}{Zhe Wang}, \bibinfo{person}{Lejian Ren}, \bibinfo{person}{Ying Fan}, \bibinfo{person}{Xiaoqiang Zhu}, {and} \bibinfo{person}{Kun Gai}.} \bibinfo{year}{2020}\natexlab{}.
\newblock \showarticletitle{Search-based user interest modeling with lifelong sequential behavior data for click-through rate prediction}. In \bibinfo{booktitle}{\emph{Proceedings of the 29th ACM International Conference on Information \& Knowledge Management}}. \bibinfo{pages}{2685--2692}.
\newblock


\bibitem[Press et~al\mbox{.}(2021)]%
        {press2021train}
\bibfield{author}{\bibinfo{person}{Ofir Press}, \bibinfo{person}{Noah~A Smith}, {and} \bibinfo{person}{Mike Lewis}.} \bibinfo{year}{2021}\natexlab{}.
\newblock \showarticletitle{Train short, test long: Attention with linear biases enables input length extrapolation}.
\newblock \bibinfo{journal}{\emph{arXiv preprint arXiv:2108.12409}} (\bibinfo{year}{2021}).
\newblock


\bibitem[Qin et~al\mbox{.}(2020)]%
        {qin2020user}
\bibfield{author}{\bibinfo{person}{Jiarui Qin}, \bibinfo{person}{Weinan Zhang}, \bibinfo{person}{Xin Wu}, \bibinfo{person}{Jiarui Jin}, \bibinfo{person}{Yuchen Fang}, {and} \bibinfo{person}{Yong Yu}.} \bibinfo{year}{2020}\natexlab{}.
\newblock \showarticletitle{User behavior retrieval for click-through rate prediction}. In \bibinfo{booktitle}{\emph{Proceedings of the 43rd International ACM SIGIR Conference on Research and Development in Information Retrieval}}. \bibinfo{pages}{2347--2356}.
\newblock


\bibitem[Qu et~al\mbox{.}(2016)]%
        {qu2016product}
\bibfield{author}{\bibinfo{person}{Yanru Qu}, \bibinfo{person}{Han Cai}, \bibinfo{person}{Kan Ren}, \bibinfo{person}{Weinan Zhang}, \bibinfo{person}{Yong Yu}, \bibinfo{person}{Ying Wen}, {and} \bibinfo{person}{Jun Wang}.} \bibinfo{year}{2016}\natexlab{}.
\newblock \showarticletitle{Product-based neural networks for user response prediction}. In \bibinfo{booktitle}{\emph{2016 IEEE 16th international conference on data mining (ICDM)}}. IEEE, \bibinfo{pages}{1149--1154}.
\newblock


\bibitem[Ren et~al\mbox{.}(2024)]%
        {ren2024representation}
\bibfield{author}{\bibinfo{person}{Xubin Ren}, \bibinfo{person}{Wei Wei}, \bibinfo{person}{Lianghao Xia}, \bibinfo{person}{Lixin Su}, \bibinfo{person}{Suqi Cheng}, \bibinfo{person}{Junfeng Wang}, \bibinfo{person}{Dawei Yin}, {and} \bibinfo{person}{Chao Huang}.} \bibinfo{year}{2024}\natexlab{}.
\newblock \showarticletitle{Representation learning with large language models for recommendation}. In \bibinfo{booktitle}{\emph{Proceedings of the ACM on Web Conference 2024}}. \bibinfo{pages}{3464--3475}.
\newblock


\bibitem[Su et~al\mbox{.}(2024)]%
        {su2024roformer}
\bibfield{author}{\bibinfo{person}{Jianlin Su}, \bibinfo{person}{Murtadha Ahmed}, \bibinfo{person}{Yu Lu}, \bibinfo{person}{Shengfeng Pan}, \bibinfo{person}{Wen Bo}, {and} \bibinfo{person}{Yunfeng Liu}.} \bibinfo{year}{2024}\natexlab{}.
\newblock \showarticletitle{Roformer: Enhanced transformer with rotary position embedding}.
\newblock \bibinfo{journal}{\emph{Neurocomputing}}  \bibinfo{volume}{568} (\bibinfo{year}{2024}), \bibinfo{pages}{127063}.
\newblock


\bibitem[Team et~al\mbox{.}(2024)]%
        {team2024gemma}
\bibfield{author}{\bibinfo{person}{Gemma Team}, \bibinfo{person}{Morgane Riviere}, \bibinfo{person}{Shreya Pathak}, \bibinfo{person}{Pier~Giuseppe Sessa}, \bibinfo{person}{Cassidy Hardin}, \bibinfo{person}{Surya Bhupatiraju}, \bibinfo{person}{L{\'e}onard Hussenot}, \bibinfo{person}{Thomas Mesnard}, \bibinfo{person}{Bobak Shahriari}, \bibinfo{person}{Alexandre Ram{\'e}}, {et~al\mbox{.}}} \bibinfo{year}{2024}\natexlab{}.
\newblock \showarticletitle{Gemma 2: Improving open language models at a practical size}.
\newblock \bibinfo{journal}{\emph{arXiv preprint arXiv:2408.00118}} (\bibinfo{year}{2024}).
\newblock


\bibitem[Wang et~al\mbox{.}(2024)]%
        {wang2024large}
\bibfield{author}{\bibinfo{person}{Jianling Wang}, \bibinfo{person}{Haokai Lu}, \bibinfo{person}{James Caverlee}, \bibinfo{person}{Ed~H Chi}, {and} \bibinfo{person}{Minmin Chen}.} \bibinfo{year}{2024}\natexlab{}.
\newblock \showarticletitle{Large Language Models as Data Augmenters for Cold-Start Item Recommendation}. In \bibinfo{booktitle}{\emph{Companion Proceedings of the ACM on Web Conference 2024}}. \bibinfo{pages}{726--729}.
\newblock


\bibitem[Wang et~al\mbox{.}(2021)]%
        {wang2021dcn}
\bibfield{author}{\bibinfo{person}{Ruoxi Wang}, \bibinfo{person}{Rakesh Shivanna}, \bibinfo{person}{Derek Cheng}, \bibinfo{person}{Sagar Jain}, \bibinfo{person}{Dong Lin}, \bibinfo{person}{Lichan Hong}, {and} \bibinfo{person}{Ed Chi}.} \bibinfo{year}{2021}\natexlab{}.
\newblock \showarticletitle{Dcn v2: Improved deep \& cross network and practical lessons for web-scale learning to rank systems}. In \bibinfo{booktitle}{\emph{Proceedings of the web conference 2021}}. \bibinfo{pages}{1785--1797}.
\newblock


\bibitem[Waswani et~al\mbox{.}(2017)]%
        {waswani2017attention}
\bibfield{author}{\bibinfo{person}{A Waswani}, \bibinfo{person}{N Shazeer}, \bibinfo{person}{N Parmar}, \bibinfo{person}{J Uszkoreit}, \bibinfo{person}{L Jones}, \bibinfo{person}{A Gomez}, \bibinfo{person}{L Kaiser}, {and} \bibinfo{person}{I Polosukhin}.} \bibinfo{year}{2017}\natexlab{}.
\newblock \showarticletitle{Attention is all you need}. In \bibinfo{booktitle}{\emph{NIPS}}.
\newblock


\bibitem[Wei et~al\mbox{.}(2023)]%
        {wei2023larger}
\bibfield{author}{\bibinfo{person}{Jerry Wei}, \bibinfo{person}{Jason Wei}, \bibinfo{person}{Yi Tay}, \bibinfo{person}{Dustin Tran}, \bibinfo{person}{Albert Webson}, \bibinfo{person}{Yifeng Lu}, \bibinfo{person}{Xinyun Chen}, \bibinfo{person}{Hanxiao Liu}, \bibinfo{person}{Da Huang}, \bibinfo{person}{Denny Zhou}, {et~al\mbox{.}}} \bibinfo{year}{2023}\natexlab{}.
\newblock \showarticletitle{Larger language models do in-context learning differently}.
\newblock \bibinfo{journal}{\emph{arXiv preprint arXiv:2303.03846}} (\bibinfo{year}{2023}).
\newblock


\bibitem[Wu et~al\mbox{.}(2024)]%
        {wu2024coral}
\bibfield{author}{\bibinfo{person}{Junda Wu}, \bibinfo{person}{Cheng-Chun Chang}, \bibinfo{person}{Tong Yu}, \bibinfo{person}{Zhankui He}, \bibinfo{person}{Jianing Wang}, \bibinfo{person}{Yupeng Hou}, {and} \bibinfo{person}{Julian McAuley}.} \bibinfo{year}{2024}\natexlab{}.
\newblock \showarticletitle{Coral: collaborative retrieval-augmented large language models improve long-tail recommendation}. In \bibinfo{booktitle}{\emph{Proceedings of the 30th ACM SIGKDD Conference on Knowledge Discovery and Data Mining}}. \bibinfo{pages}{3391--3401}.
\newblock


\bibitem[Xi et~al\mbox{.}(2024)]%
        {xi2024towards}
\bibfield{author}{\bibinfo{person}{Yunjia Xi}, \bibinfo{person}{Weiwen Liu}, \bibinfo{person}{Jianghao Lin}, \bibinfo{person}{Xiaoling Cai}, \bibinfo{person}{Hong Zhu}, \bibinfo{person}{Jieming Zhu}, \bibinfo{person}{Bo Chen}, \bibinfo{person}{Ruiming Tang}, \bibinfo{person}{Weinan Zhang}, {and} \bibinfo{person}{Yong Yu}.} \bibinfo{year}{2024}\natexlab{}.
\newblock \showarticletitle{Towards open-world recommendation with knowledge augmentation from large language models}. In \bibinfo{booktitle}{\emph{Proceedings of the 18th ACM Conference on Recommender Systems}}. \bibinfo{pages}{12--22}.
\newblock


\bibitem[Yang et~al\mbox{.}(2024)]%
        {yang2024qwen2}
\bibfield{author}{\bibinfo{person}{An Yang}, \bibinfo{person}{Baosong Yang}, \bibinfo{person}{Beichen Zhang}, \bibinfo{person}{Binyuan Hui}, \bibinfo{person}{Bo Zheng}, \bibinfo{person}{Bowen Yu}, \bibinfo{person}{Chengyuan Li}, \bibinfo{person}{Dayiheng Liu}, \bibinfo{person}{Fei Huang}, \bibinfo{person}{Haoran Wei}, {et~al\mbox{.}}} \bibinfo{year}{2024}\natexlab{}.
\newblock \showarticletitle{Qwen2. 5 Technical Report}.
\newblock \bibinfo{journal}{\emph{arXiv preprint arXiv:2412.15115}} (\bibinfo{year}{2024}).
\newblock


\bibitem[Yang et~al\mbox{.}(2023)]%
        {yang2023palr}
\bibfield{author}{\bibinfo{person}{Fan Yang}, \bibinfo{person}{Zheng Chen}, \bibinfo{person}{Ziyan Jiang}, \bibinfo{person}{Eunah Cho}, \bibinfo{person}{Xiaojiang Huang}, {and} \bibinfo{person}{Yanbin Lu}.} \bibinfo{year}{2023}\natexlab{}.
\newblock \showarticletitle{Palr: Personalization aware llms for recommendation}.
\newblock \bibinfo{journal}{\emph{arXiv preprint arXiv:2305.07622}} (\bibinfo{year}{2023}).
\newblock


\bibitem[Zhai et~al\mbox{.}(2024)]%
        {zhai2024actions}
\bibfield{author}{\bibinfo{person}{Jiaqi Zhai}, \bibinfo{person}{Lucy Liao}, \bibinfo{person}{Xing Liu}, \bibinfo{person}{Yueming Wang}, \bibinfo{person}{Rui Li}, \bibinfo{person}{Xuan Cao}, \bibinfo{person}{Leon Gao}, \bibinfo{person}{Zhaojie Gong}, \bibinfo{person}{Fangda Gu}, \bibinfo{person}{Michael He}, {et~al\mbox{.}}} \bibinfo{year}{2024}\natexlab{}.
\newblock \showarticletitle{Actions speak louder than words: Trillion-parameter sequential transducers for generative recommendations}.
\newblock \bibinfo{journal}{\emph{arXiv preprint arXiv:2402.17152}} (\bibinfo{year}{2024}).
\newblock


\bibitem[Zhang et~al\mbox{.}(2023c)]%
        {zhang2023recommendation}
\bibfield{author}{\bibinfo{person}{Junjie Zhang}, \bibinfo{person}{Ruobing Xie}, \bibinfo{person}{Yupeng Hou}, \bibinfo{person}{Wayne~Xin Zhao}, \bibinfo{person}{Leyu Lin}, {and} \bibinfo{person}{Ji-Rong Wen}.} \bibinfo{year}{2023}\natexlab{c}.
\newblock \showarticletitle{Recommendation as instruction following: A large language model empowered recommendation approach}.
\newblock \bibinfo{journal}{\emph{arXiv preprint arXiv:2305.07001}} (\bibinfo{year}{2023}).
\newblock


\bibitem[Zhang et~al\mbox{.}(2023a)]%
        {zhang2023collm}
\bibfield{author}{\bibinfo{person}{Yang Zhang}, \bibinfo{person}{Fuli Feng}, \bibinfo{person}{Jizhi Zhang}, \bibinfo{person}{Keqin Bao}, \bibinfo{person}{Qifan Wang}, {and} \bibinfo{person}{Xiangnan He}.} \bibinfo{year}{2023}\natexlab{a}.
\newblock \showarticletitle{Collm: Integrating collaborative embeddings into large language models for recommendation}.
\newblock \bibinfo{journal}{\emph{arXiv preprint arXiv:2310.19488}} (\bibinfo{year}{2023}).
\newblock


\bibitem[Zhang et~al\mbox{.}(2023b)]%
        {collm}
\bibfield{author}{\bibinfo{person}{Yang Zhang}, \bibinfo{person}{Fuli Feng}, \bibinfo{person}{Jizhi Zhang}, \bibinfo{person}{Keqin Bao}, \bibinfo{person}{Qifan Wang}, {and} \bibinfo{person}{Xiangnan He}.} \bibinfo{year}{2023}\natexlab{b}.
\newblock \showarticletitle{Collm: Integrating collaborative embeddings into large language models for recommendation}.
\newblock \bibinfo{journal}{\emph{arXiv preprint arXiv:2310.19488}} (\bibinfo{year}{2023}).
\newblock


\bibitem[Zhao et~al\mbox{.}(2023)]%
        {zhao2023survey}
\bibfield{author}{\bibinfo{person}{Wayne~Xin Zhao}, \bibinfo{person}{Kun Zhou}, \bibinfo{person}{Junyi Li}, \bibinfo{person}{Tianyi Tang}, \bibinfo{person}{Xiaolei Wang}, \bibinfo{person}{Yupeng Hou}, \bibinfo{person}{Yingqian Min}, \bibinfo{person}{Beichen Zhang}, \bibinfo{person}{Junjie Zhang}, \bibinfo{person}{Zican Dong}, {et~al\mbox{.}}} \bibinfo{year}{2023}\natexlab{}.
\newblock \showarticletitle{A survey of large language models}.
\newblock \bibinfo{journal}{\emph{arXiv preprint arXiv:2303.18223}} (\bibinfo{year}{2023}).
\newblock


\bibitem[Zheng et~al\mbox{.}(2024)]%
        {zheng2024adapting}
\bibfield{author}{\bibinfo{person}{Bowen Zheng}, \bibinfo{person}{Yupeng Hou}, \bibinfo{person}{Hongyu Lu}, \bibinfo{person}{Yu Chen}, \bibinfo{person}{Wayne~Xin Zhao}, \bibinfo{person}{Ming Chen}, {and} \bibinfo{person}{Ji-Rong Wen}.} \bibinfo{year}{2024}\natexlab{}.
\newblock \showarticletitle{Adapting large language models by integrating collaborative semantics for recommendation}. In \bibinfo{booktitle}{\emph{2024 IEEE 40th International Conference on Data Engineering (ICDE)}}. IEEE, \bibinfo{pages}{1435--1448}.
\newblock


\bibitem[Zhou et~al\mbox{.}(2018)]%
        {zhou2018deep}
\bibfield{author}{\bibinfo{person}{Guorui Zhou}, \bibinfo{person}{Xiaoqiang Zhu}, \bibinfo{person}{Chenru Song}, \bibinfo{person}{Ying Fan}, \bibinfo{person}{Han Zhu}, \bibinfo{person}{Xiao Ma}, \bibinfo{person}{Yanghui Yan}, \bibinfo{person}{Junqi Jin}, \bibinfo{person}{Han Li}, {and} \bibinfo{person}{Kun Gai}.} \bibinfo{year}{2018}\natexlab{}.
\newblock \showarticletitle{Deep interest network for click-through rate prediction}. In \bibinfo{booktitle}{\emph{Proceedings of the 24th ACM SIGKDD international conference on knowledge discovery \& data mining}}. \bibinfo{pages}{1059--1068}.
\newblock


\bibitem[Zhu et~al\mbox{.}(2024)]%
        {zhu2024collaborative}
\bibfield{author}{\bibinfo{person}{Yaochen Zhu}, \bibinfo{person}{Liang Wu}, \bibinfo{person}{Qi Guo}, \bibinfo{person}{Liangjie Hong}, {and} \bibinfo{person}{Jundong Li}.} \bibinfo{year}{2024}\natexlab{}.
\newblock \showarticletitle{Collaborative large language model for recommender systems}. In \bibinfo{booktitle}{\emph{Proceedings of the ACM on Web Conference 2024}}. \bibinfo{pages}{3162--3172}.
\newblock


\end{thebibliography}

\clearpage
\section{Appendix}

\noindent\textbf{Implementation Details.} We select Llama-3.1-8B-Instruct \cite{dubey2024llama} released by Meta as the base LLM for experimentation. All experiments are conducted using A100 GPUs. For resource efficiency, 16-bit quantization and low-rank adaption (LoRA) \cite{hu2021lora} are adopted for parameter-efficient finetuning (PEFT). We select the LoRA rank from $\{8, 16\}$, alpha from $\{16, 32\}$, and the dropout rate from $\{0.05, 0.1\}$. We tune the following LoRA projection matrices: query, key, value, output, up, down, and gate. During fine-tuning, we adopt the AdamW optimizer \cite{kingma2014adam} with the weight decay set to $0.001$. The LLM is trained with a batch size of $64$ and the learning rate is chosen from $\{1e-5, 2e-5, 5e-5, 1e-4\}$ with a cosine scheduler and a warm-up ratio of $0.1$. For all three data sets, the maximum training epoch is set to 10 with an early stopping patience set to 2. To construct the textual input prompt, we follow existing work \cite{lin2024rella, bao2023tallrec} to remove all pure ID fields from both MovieLens and Kindles, as LLM possess limited perceptual abilities for pure ID texts \cite{lin2023can}. For MovieLens 1M, we use all available item features -- title and genre -- to describe each interacted item. For CDs and Kindles, instead of using all available item features, we only use title, category, and brand to avoid overly lengthy prompts. For all three datasets, we use a consistent context window consisting of $20$ interactions. Empirically, since each interaction is represented by its tokenized textual descriptions, the windowed causal attention mechanism is equivalent to applying a sliding window attention to allow each token to only attend to its previous $N$ tokens, where $N$ denotes the number of tokens associated with the previous $n$ context interactions. However, since we do not explicitly pad or truncate all interactions to be represented by the same amount of tokens, the value of $N$ varies from item to item. To further increase the computational efficiency, we fix the context interaction window size for all target interactions to be $1024$.

\smallskip
\noindent\textbf{Sample Sliding Window Prompt.}
Below we present a sample prompt constructed under the conventional sliding-window approach. Each interaction is being represented by its textual descriptions to make the prompt comprehensible to the LLM. Each prompt starts with a message instructing the LLM to predict the user's preference on the target interaction based on his or her most recent $n$ interaction history, which is provided right after the starting instruction.
\begin{figure}[H]
  \includegraphics[width=.5\textwidth]{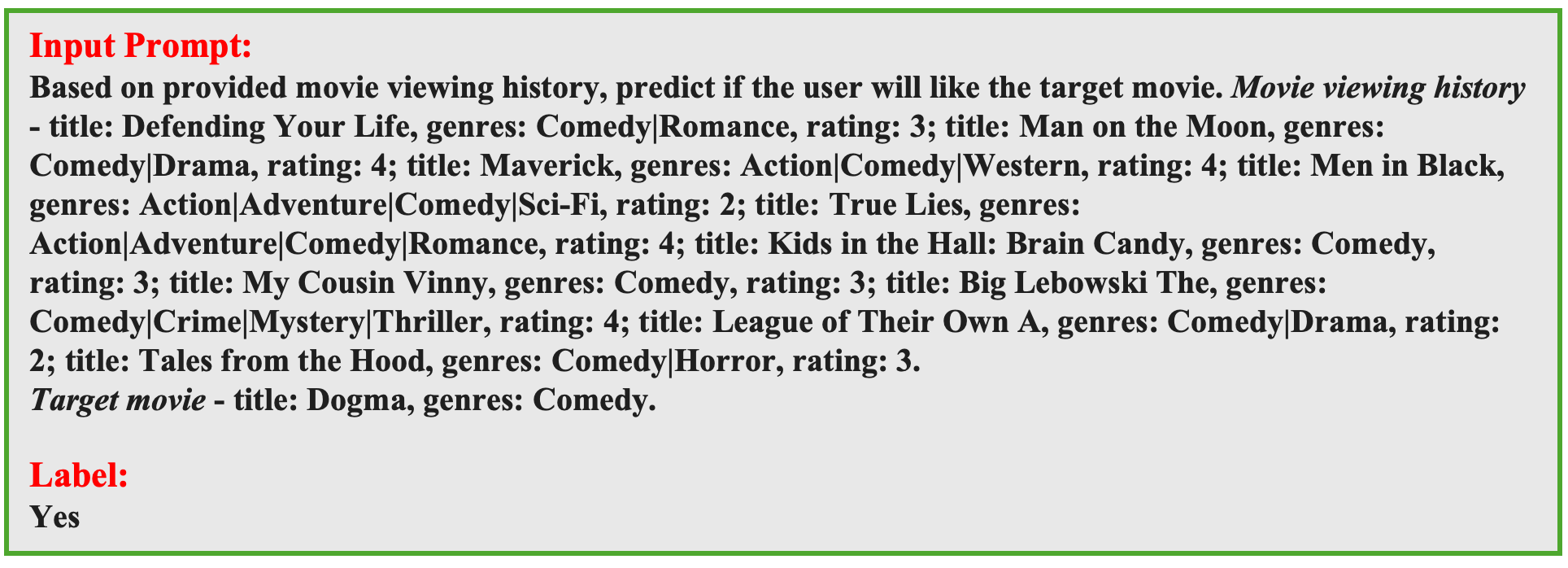}
  \caption{Illustration of a sliding-window prompt with its corresponding label}
  \label{fig:prompt}
\end{figure}

\begin{figure}[H]
  \includegraphics[width=.5\textwidth]{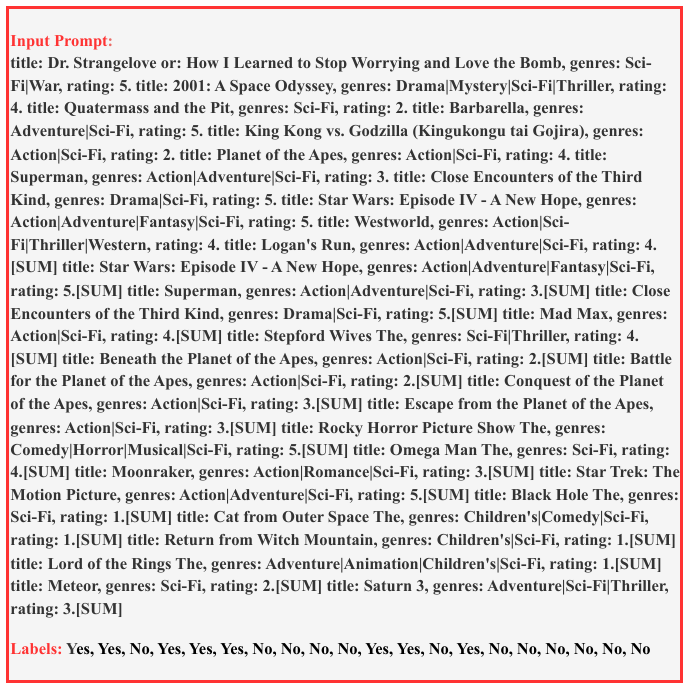}
  \caption{Illustration of a streaming prompt with its corresponding labels}
  \label{fig:stream_prompt}
\end{figure}

\smallskip
\noindent\textbf{Sample Streaming Prompt.}
Below we present a sample prompt under the streaming prompt formulation that DTI employs. Each interaction is being represented by its textual descriptions, same as in the sliding-window approach. Each interaction after the first $n$ context interaction is being followed by a special [SUM] token to be used to compute the language-modeling cross entropy loss with the target interaction's corresponding label. It is evident in \Cref{fig:stream_prompt} that the streaming prompt formulation results in substantially longer prompts compared to the sliding-window prompt formulation. Therefore, the integration of the windowed casual attention mechanism becomes vital to reduce the quadratic self-attention scaling with the increased prompt length to a linear scaling.

\end{document}